\documentclass[twocolumn]{article}
\usepackage{indentfirst,amsmath,latexsym,amssymb}
\newcommand{\be}{\begin{equation}}
\newcommand{\ee}{\end{equation}}
\newcommand{\ra}{\rightarrow}
\newcommand{\lra}{\leftrightarrow}
\newcommand{\qed}{\hfill $\bullet$}

\newtheorem{prop}{Proposition}
\newtheorem{thm}[prop]{Theorem}
\newtheorem{lem}[prop]{Lemma}
\newtheorem{cor}[prop]{Corollary}
\newenvironment{defn}{\trivlist \item[\hskip \labelsep{\bf Definition.}]}{\qed \endtrivlist}
\newenvironment{prf}{\trivlist \item[\hskip \labelsep{\bf Proof.}]}{\qed \endtrivlist}

\begin{document}
\newcommand{\bldc}{{\mathbf{c}}}
\newcommand{\bldp}{{\mathbf{p}}}
\newcommand{\bldg}{{\mathbf{g}}}
\newcommand{\tldcf}{{\tilde{F}}}
\newcommand{\hatb}{{\hat{b}}}
\newcommand{\tldb}{{\tilde{b}}}
\newcommand{\hatx}{{\hat{x}}}
\newcommand{\hatf}{{\hat{f}}}
\newcommand{\bldhatg}{{\hat{\mathbf{g}}}}
\newcommand{\bldhate}{{\hat{\mathbf{e}}}}
\newcommand{\ifof}{if and only if}
\newcommand{\bone}{\mathbf{1}}
\newcommand{\sm}{Schreier matrix}
\newcommand{\csm}{controllable Schreier matrix}
\newcommand{\dsm}{dual Schreier matrix}
\newcommand{\xj}{{\{X_j\}}}
\newcommand{\yk}{{\{Y_k\}}}

\newcommand{\calg}{{\mathcal{G}}}
\newcommand{\tldcalg}{{\tilde{\mathcal{G}}}}
\newcommand{\calv}{{\mathcal{V}}}
\newcommand{\cale}{{\mathcal{E}}}
\newcommand{\lctl}{$l$-controllable}
\newcommand{\ellctl}{$\ell$-controllable}

\title{Group codes and the Schreier matrix form}
\author{Kenneth M. Mackenthun Jr.}
\maketitle

\vspace{3mm}
{\bf ABSTRACT}
\vspace{3mm}

In a group trellis, the sequence of branches that split from the 
identity path and merge to the identity path form two normal chains.
The Schreier refinement theorem can be applied to these two normal
chains.  The refinement of the two normal chains can be written in 
the form of a matrix, called the \sm\ form, with rows and columns 
determined by the two normal chains.

Based on the \sm\ form,
we give an encoder structure for a group code which is an estimator.
The encoder uses the important idea of shortest length generator 
sequences previously explained by Forney and Trott.  In this encoder
the generator sequences are shown to have an additional property:
the components of the generators are coset representatives in
a chain coset decomposition of the branch group $B$ of the code.
Therefore this encoder appears to be a natural form 
for a group code encoder.  The encoder has a register implementation 
which is somewhat different from the classical shift register structure.

This form of the encoder can be extended.  We find a composition chain
of the branch group $B$ and give an encoder which uses coset representatives
in the composition chain of $B$.  When $B$ is solvable, the generators
are constructed using coset representatives taken from prime cyclic groups.

\vspace{3mm}
{\bf 1.  INTRODUCTION}
\vspace{3mm}

The idea of group codes and group shifts is important in several areas
of mathematics and engineering such as symbolic dynamics,
linear systems theory, and coding theory.  Some of the seminal 
papers in these areas are the work of Kitchens \cite{KIT}, Willems \cite{JW},
Forney and Trott \cite{FT}, and Loeliger and Mittelholzer \cite{LM}.

Kitchens \cite{KIT} introduced the idea of a group shift \cite{LMr}
and showed that a group shift has finite memory, i.e., 
it is a shift of finite type \cite{LMr}.
A group shift is a fundamental example of a time invariant group code \cite{FT}.
Forney and Trott \cite{FT} show that any time invariant group code is
equivalent to a labeled group trellis section.  
They show that any group code that is complete
(any global constraints can be determined locally, cf. \cite{FT}) is
equivalent to a sequence of connected labeled group trellis sections (which may
vary in time).  They explain the important idea of shortest length code
sequences, or generator sequences.  They show that at each time epoch,
a finite set of generator sequences can be used construct a ``local"
section of the code.  Using the generator sequences,
they show that any group code can be mechanized
with a minimal encoder which has a shift register structure.

Forney and Trott use a ``top down" approach, that is, they start with a
group code, a set of sequences with a group property, and then analyze
further to determine the state structure, encoder structure, and other
properties of the code; the related work of \cite{JW} in systems theory 
also uses a top down approach.  The work of Loeliger and Mittelholzer 
\cite{LM} is a ``bottom up" approach.  They start with a group trellis
section, and use it to construct a group trellis, or group code.
Among other results, Loeliger and Mittelholzer give an
abstract characterization of groups which can be the
branch group $B$ of a group code.  They also give a shift register structure
for a group trellis.  The development of their encoder using a bottom up
approach is in some sense a mirror image of the Forney and Trott top 
down encoder construction.

In this paper, we use the bottom up approach of Loeliger and Mittelholzer. 
We start with a group trellis section and determine properties of the
group trellis.

In a group trellis, the sequence of branches that split from the 
identity path and merge to the identity path form two normal chains,
$\xj$ and $\yk$, respectively.
These two normal chains were first used in the work of \cite{LM}.
In this paper we apply the Schreier refinement theorem to $\xj$ and $\yk$. 
The refinement of the two normal chains can be written in 
the form of a matrix, called the \sm\ form, with rows and columns 
determined by $\xj$ and $\yk$.  When the group trellis is 
controllable, the \sm\ is a triangular form.  The \sm\ is an echo of
matrix ideas used in classical linear systems analysis.

Based on the \sm\ form,
we give an encoder structure for a group code which is an estimator.
Both encoders of \cite{FT,LM} use shortest length generator sequences.
The encoder here also uses shortest length generator sequences, but in this encoder
the generator sequences are shown to have an additional property:
the components of the generators are coset representatives in
a chain coset decomposition of the branch group $B$ of the code.
This shows that the generator code sequences are intimately related
to the structure of the branch group.
Therefore this encoder appears to be a natural form 
for a group code encoder.  In addition, the encoder has a convolution
property which is not apparent in the encoders of \cite{FT,LM}.
The encoder has a register implementation 
which is somewhat different from the classical shift register structure.

This approach can be iterated.  We use properties of the
group trellis to find a composition chain of $B$.  We insert this 
composition chain into $\xj$ to find a refinement of the \sm\
which is a composition chain, a Schreier array.
Using the Schreier array, we give an encoder which uses coset representatives
in the composition chain of $B$.  When $B$ is solvable, the generators
are constructed using coset representatives taken from prime cyclic groups.

This paper is organized as follows.  Section 2 defines a group trellis
section and group trellis.  We study an \ellctl\ group trellis,
in which each state can be reached from any other state in $\ell$ branches.
Section 3 defines the \sm\ and a controllable \sm, which has a triangular form.
In Section 4, we analyze the structure of the controllable \sm\ and the controllable
group trellis.  We focus on the branches that split from the identity path,
the sequence $\xj$.  The only technical tool we use is several
simple generalizations of the correspondence theorem.  Based on the 
analysis in Section 4, Section 5 gives an encoder for the group trellis.
The encoder uses shortest length generator sequences and has a 
convolution property.  The shortest length generator sequences are 
composed of coset representatives from a chain coset decomposition of $B$.
In Section 6, we use properties of the group trellis to give a 
composition series of $B$.  In Section 7, we use the composition series of
$B$ and the mechanics of the \sm\ to give a novel encoder for the group 
trellis.  The encoder uses shortest length generator sequences whose
components are coset representatives in the composition series of $B$.

\vspace{3mm}
{\bf 2.  GROUP TRELLIS}
\vspace{3mm}

As in \cite{MH}, we say $G$ is a {\it subdirect product}
of $G_1$ and $G_2$ if it is a subgroup of $G_1\times G_2$ and the
first and second coordinate of $G$ take all values in $G_1$ and
$G_2$, respectively; we also say $G$ is a subdirect product of
$G_1\times G_2$.

\begin{defn}
A {\it group trellis section} is a subdirect product $B$ of $S\times S$.
(Therefore the left and right coordinates of $B$ use all elements of $S$.)
We call $B$ the {\it branch group} and $S$ the {\it state group}.
The elements of $B$ are {\it branches} $b$, where $b=(s,s')\in S\times S$.
\end{defn}

We can think of a group trellis section $B$ as a bipartite graph
with branches in $B$ and vertices in $S$, where there is a branch $(s,s')$ 
between two vertices $s$ and $s'$ \ifof\ $(s,s')\in B$.  

\begin{defn}  
A {\it labeled group trellis section} is a 
subdirect product $\tilde{B}$ of $S\times A\times S$.
We call $A$ the {\it label group} or {\it alphabet}.
The elements of $\tilde{B}$ are {\it labeled branches} $\tldb$, 
where $\tldb=(s,a,s')\in S\times A\times S$.
\end{defn}

Note that there is a homomorphism $\omega:  B\ra A$.
Note that if $A=\bone$, then the labeled group trellis $\tilde{B}$ is isomorphic to 
the group trellis $B$.

\begin{defn}
A {\it group trellis} $C$ is the shift of a group trellis section.
A {\it labeled group trellis} $\tilde{C}$ is the shift of a 
labeled group trellis section.
\end {defn}

We only consider group codes defined on the integers $\mathbf{Z}$.

\begin{defn}
A {\it group code} $\mathcal{C}$ is a subgroup of an infinite direct product
group $\prod_{\mathbf{Z}} A$, where $\mathbf{Z}$ is the integers.
\end {defn}

Consider the projection map $\pi^-:  B\ra S$ given by the assignment
$(s,s')\mapsto s$.  This is a homomorphism with kernel $X_0$,
the subgroup of all elements of $B$ of the form $(\bone,s')$.  Let $B^-$ be
the left states of $B$, so that $B^-=S$.  Then
$$
\frac{B}{X_0}\simeq B^-=S.
$$
Consider the projection map $\pi^+:  B\ra S$ given by the assignment
$(s,s')\mapsto s'$.  This is a homomorphism with kernel $Y_0$,
the subgroup of all elements of $B$ of the form $(s,\bone)$.  Let $B^+$ be
the right states of $B$, so that $B^+=S$.  Then
$$
\frac{B}{Y_0}\simeq B^+=S.
$$
Together these give 
$$
\frac{B}{X_0}\simeq B^-=S=B^+\simeq\frac{B}{Y_0}.
$$

\begin{prop}
\label{prop1a}
Let $G$ be a subgroup of $B$.  We have $G\cap X_0\lhd G$ and
$G\cap Y_0\lhd G$.  The branches that split from each left state of $G$, 
$G^-$, are a coset of $G\cap X_0$.  The branches that merge to each 
right state of $G$, $G^+$, are a coset of $G\cap Y_0$.
\end{prop}

We think of a group trellis $C$ as a connected sequence of group trellis
sections $B$.  The states of the group trellis occur at time epochs
$t$, which are integers in the range $-\infty<t<\infty$.  The states at time
epoch $t$ are $S_t$, where $S_t=S$.  The trellis section $B_t=B$
is the trellis section with left states at time epoch $t$, $S_t$,
and right states at time epoch $t+1$, $S_{t+1}$.  The right states
of $B_t$, $S_{t+1}$, are the left states of $B_{t+1}$.  
By the phrase {\it a branch at time epoch $t$} we mean a branch $b_t\in B_t$;
we have $b_t=(s,s')$ where $s\in S_t$ and $s'\in S_{t+1}$.

Let $\bldc$ be a trellis path in $C$.  Then 
$$
\bldc=\ldots,b_{-1},b_0,b_1,\ldots,b_t,\ldots,
$$
where component $b_t$ is a branch in $B_t=B$.  Define the projection map
$\chi_t:  C\ra B_t$ by the assignment $\bldc\mapsto b_t$.  In
other words $\chi_t(\bldc)=b_t=(s,s')$ where $s\in B_t$ and $s'\in B_{t+1}$.
Define the projection map $\chi_{[t_1,t_2]}:  C\ra B_{t_1}\times\cdots\times B_{t_2}$ 
by the assignment $\bldc\mapsto (b_{t_1},\ldots,b_{t_2})$, the components
of $\bldc$ over the time interval $[t_1,t_2]$.  For all integers $i$,
define $C_{[i,\infty)}$ to
be the set of paths $\bldc$ such that components $b_t=\bone$ for $t<i$, where
$\bone$ is the identity of $B$.  Then $C_{[i,\infty)}$ is the set of paths
that are in the identity state at time epochs in the interval $(-\infty,i]$.
For all integers $i$, define $C_{(-\infty,i]}$ to
be the set of paths $\bldc$ such that components $b_t=\bone$ for $t>i$.
For all integers $i$, let
$$
X_i=\{\chi_0(\bldc) | \bldc\in C_{[-i,\infty)}\}.
$$
Note that $X_i=\bone$ for $i<0$.
For all integers $i$, let
$$
Y_i=\{\chi_0(\bldc) | \bldc\in C_{(-\infty,i]}\}.
$$
Note that $Y_i=\bone$ for $i<0$.  As defined, $X_i$ and $Y_i$ are
elements of $B$; a priori there is no intrinsic notion of time 
associated with subscript $i$.  Then for each $i$, $X_i\in B_t$ and 
$Y_i\in B_t$ for all time epochs $t$, $-\infty<t<\infty$.
Note that $X_i^+=X_{i+1}^-$ and $Y_i^+=Y_{i-1}^-$ for all integers $i$.  And
$X_i\lhd B$ and $Y_i\lhd B$ for all integers $i$ \cite{LM}.

We say a group trellis $C$ is {\it $l$-controllable} if for any time 
epoch $t$, and any pair of states $(s,s')\in S_t\times S_{t+l}$, 
there is a trellis path through these two states.  The least integer $l$ 
for which a group trellis is \lctl\ is denoted as $\ell$.  
In this paper, we only study the case $l=\ell$.

The next result follows directly from Proposition 7.2 of \cite{LM},
using our notation.

\begin{prop}
\label{prop1}
The group trellis $C$ is $\ell$-controllable \ifof\
$(X_{\ell-1})^+=B^-$, or equivalently, \ifof\ $(Y_{\ell-1})^-=B^+$.
\end{prop}

\vspace{3mm}
{\bf 3.  SCHREIER MATRIX FORM}
\vspace{3mm}

The group $B$ has two normal series (and chief series)
$$ \bone=X_{-1}\subset X_0\subset X_1 \subset \cdots \subset X_\ell=B  $$
and
$$ \bone=Y_{-1}\subset Y_0\subset Y_1 \subset \cdots \subset Y_\ell=B.  $$
We denote these normal series by $\xj$ and $\yk$.  The Schreier
refinement theorem used to prove the Jordan-H\"{o}lder theorem \cite{ROT} shows
how to obtain a refinement of $\xj$ by inserting $\yk$, as shown
in equation (\ref{sm}) (see next page).  In (\ref{sm}), we have written
the refinement as a matrix of $\ell+1$ columns and $\ell+2$ rows.
Note that the terms in the bottom row form the
sequence $X_{-1},X_0,X_1, \ldots X_{\ell-2},X_{\ell-1}$,
and the terms in the top row form the sequence
$X_0,X_1,X_2, \ldots X_{\ell-1},X_\ell$.
Thus (\ref{sm}) is indeed a refinement of the normal series $\xj$.
We call (\ref{sm}) the {\it Schreier matrix form of $\xj$ and $\yk$},
or (loosely) just the \sm\ of $\xj$ and $\yk$.
Since $\xj$ and $\yk$ are chief series, the \sm\ of $\xj$ and $\yk$ is a chief series.

The {\it diagonal terms} of the Schreier matrix are $X_{j-1}(X_j\cap Y_{\ell-j})$ for
$j=0,\ldots\ell$.  We say the Schreier matrix is {\it $\ell$-controllable}
if $X_{j-1}(X_j\cap Y_{\ell-j})=X_j$ for
$j=0,\ldots\ell$.  This is trivially satisfied for $j=0$.  For $j\in[1,\ell]$,
this means all column terms above the diagonal term are the same as
the diagonal term.  Then we can reduce the \sm\ and write
the \ellctl\ \sm\ as in (\ref{csm}).  Note that the \ellctl\ \sm\ is a
triangular form.  (A triangle can be formed in two ways, depending on
whether the columns in (\ref{sm}) are shifted up or not; we have shifted
the columns up since it is more useful here.)
Note that the \sm\ term $X_{j-1}(X_j\cap Y_k)$ is in the $j^{\rm th}$ column 
of (\ref{csm}), for $0\le j\le\ell$, and $(j+k)^{\rm th}$ row, for
$0\le j+k\le\ell$ (counting up from the bottom).

\begin{prop}
If the group trellis $C$ is \ellctl, then the \sm\ of $\xj$ and $\yk$
is \ellctl.
\end{prop}

\begin{prf}
If the group trellis $C$ is \ellctl, then from Proposition 7.2 of \cite{LM}
(using our notation),
\be
\label{eq111}
(X_0\cap Y_\ell)(X_1\cap Y_{\ell-1})\cdots (X_j\cap Y_{\ell-j})=X_j
\ee
for all $j\ge 0$.  This means we can rewrite (\ref{eq111}) as
$$
X_{j-1}(X_j\cap Y_{\ell-j})=X_j
$$
for all $j\ge 0$.  Then the \sm\ of $\xj$ and $\yk$ is \ellctl.
\end{prf}

The {\it Schreier matrix of $\yk$ and $\xj$} is obtained by interchanging 
$X$ and $Y$ in (\ref{sm}); it is the dual of the \sm\ of $\xj$ and $\yk$.  
We say the \sm\ of $\yk$ and $\xj$ is \ellctl\ if
$Y_{k-1}(Y_k\cap X_{\ell-k})=Y_k$ for $k=0,\ldots\ell$.

\begin{prop}
The \sm\ $\mathbf{M}$ of $\xj$ and $\yk$ is \ellctl\ \ifof\ the (dual) \sm\
$\mathbf{M}_d$ of $\yk$ and $\xj$ is \ellctl.
\end{prop}

\begin{prf}
By the Zassenhaus lemma used to prove the Schreier refinement theorem \cite{ROT},
we have
\be
\label{zlem}
\frac{X_{j-1}(X_j\cap Y_k)}{X_{j-1}(X_j\cap Y_{k-1})}\simeq
\frac{Y_{k-1}(Y_k\cap X_j)}{Y_{k-1}(Y_k\cap X_{j-1})},
\ee
for $j$ and $k$ in the range $0\le j\le\ell$, $0\le k\le\ell$.
Note that the numerator and denominator on the left are in the same column
of $\mathbf{M}$ (see (\ref{sm})) and the numerator and denominator on the right are in 
the same column of $\mathbf{M}_d$.  Assume the \sm\ is \ellctl.
If $j+k-1\ge\ell$, then the denominator term on the left is on the diagonal
or above, and the left hand side is isomorphic to $\bone$.  This means
the right hand side is isomorphic to $\bone$.  
But if $j+k-1\ge\ell$, the denominator term on the right is on the diagonal
or above, and then the \dsm\ is \ellctl.  The reverse direction 
is the same proof.
\end{prf}

\begin{figure*}[pt]
\be
\label{sm}
\begin{array}{cccccc}
  \cup & \cup & \cup && \cup & \\

  X_{-1}(X_0\cap Y_\ell) & X_0(X_1\cap Y_\ell) & X_1(X_2\cap Y_\ell) & \cdots & X_{\ell-2}(X_{\ell-1}\cap Y_\ell) & X_{\ell-1}(X_\ell\cap Y_\ell) \\

  \cup & \cup & \cup && \cup & \cup \\

  X_{-1}(X_0\cap Y_{\ell-1}) & X_0(X_1\cap Y_{\ell-1}) & X_1(X_2\cap Y_{\ell-1}) & \cdots & X_{\ell-2}(X_{\ell-1}\cap Y_{\ell-1}) & X_{\ell-1}(X_\ell\cap Y_{\ell-1}) \\

  \cup & \cup & \cup && \cup & \cup \\

  X_{-1}(X_0\cap Y_{\ell-2}) & X_0(X_1\cap Y_{\ell-2}) & X_1(X_2\cap Y_{\ell-2}) & \cdots & X_{\ell-2}(X_{\ell-1}\cap Y_{\ell-2}) & X_{\ell-1}(X_\ell\cap Y_{\ell-2}) \\

  \cup & \cup & \cup && \cup & \cup \\

  \cdots & \cdots & \cdots & \cdots & \cdots & \cdots \\

  \cup & \cup & \cup && \cup & \cup \\

  X_{-1}(X_0\cap Y_2) & X_0(X_1\cap Y_2) & X_1(X_2\cap Y_2) & \cdots & X_{\ell-2}(X_{\ell-1}\cap Y_2) & X_{\ell-1}(X_\ell\cap Y_2) \\

  \cup & \cup & \cup && \cup & \cup \\

  X_{-1}(X_0\cap Y_1) & X_0(X_1\cap Y_1) & X_1(X_2\cap Y_1) & \cdots & X_{\ell-2}(X_{\ell-1}\cap Y_1) & X_{\ell-1}(X_\ell\cap Y_1) \\

  \cup & \cup & \cup && \cup & \cup \\

  X_{-1}(X_0\cap Y_0) & X_0(X_1\cap Y_0) & X_1(X_2\cap Y_0) & \cdots & X_{\ell-2}(X_{\ell-1}\cap Y_0) & X_{\ell-1}(X_\ell\cap Y_0) \\

  \cup & \cup & \cup && \cup & \cup \\

  X_{-1}(X_0\cap Y_{-1}) & X_0(X_1\cap Y_{-1}) & X_1(X_2\cap Y_{-1}) & \cdots & X_{\ell-2}(X_{\ell-1}\cap Y_{-1}) & X_{\ell-1}(X_\ell\cap Y_{-1})
\end{array}
\ee
\end{figure*}

\begin{figure*}[pb]
\be
\label{csm}
\begin{array}{ccccccc}
  \shortparallel & \shortparallel & \shortparallel && \shortparallel & \shortparallel & \\

  X_{-1}(X_0\cap Y_\ell) & X_0(X_1\cap Y_{\ell-1}) & X_1(X_2\cap Y_{\ell-2}) & \cdots & X_{\ell-2}(X_{\ell-1}\cap Y_1) & X_{\ell-1}(X_\ell\cap Y_0) & X_\ell(\bone) \\

  \cup & \cup & \cup && \cup & \cup & \\

  X_{-1}(X_0\cap Y_{\ell-1}) & X_0(X_1\cap Y_{\ell-2}) & X_1(X_2\cap Y_{\ell-3}) & \cdots  & X_{\ell-2}(X_{\ell-1}\cap Y_0) & X_{\ell-1}(\bone) \\

  \cup & \cup & \cup && \cup && \\

  X_{-1}(X_0\cap Y_{\ell-2}) & X_0(X_1\cap Y_{\ell-3}) & X_1(X_2\cap Y_{\ell-4}) & \cdots  & X_{\ell-2}(\bone) && \\
 
  \cup & \cup & \cup &&&& \\

  \cdots & \cdots & \cdots &&&& \\

  \cup & \cup & \cup &&&& \\

  X_{-1}(X_0\cap Y_2) & X_0(X_1\cap Y_1) & X_1(X_2\cap Y_0) &&&& \\

  \cup & \cup & \cup &&&& \\

  X_{-1}(X_0\cap Y_1) & X_0(X_1\cap Y_0) & X_1(\bone) &&&& \\

  \cup & \cup &&&&& \\
   
  X_{-1}(X_0\cap Y_0) & X_0(\bone) &&&&& \\

  \cup &&&&&& \\

  X_{-1}(\bone) &&&&&&
\end{array}
\ee
\end{figure*}

\vspace{3mm}
{\bf 4.  STRUCTURE OF THE CONTROLLABLE SCHREIER MATRIX FORM}
\vspace{3mm}

We now show that terms in the \csm\ in (\ref{csm}) can be related to
certain paths in the trellis.  
Consider paths in the trellis which split from the identity state at
time epoch $0$.  The branches in 
these paths are $X_0$ at time epoch $0$, $X_1$ at time epoch $1$, $\ldots$, and
$X_\ell$ at time epoch $\ell$.  These paths are represented by the upper sloped
line in Figure \ref{fig1}, labeled $X_0,X_1,\ldots,X_\ell$.
Now consider those paths that split from the identity
at time $1$.  The branches in 
these paths are $X_0$ at time $1$, $X_1$ at time $2$, $\ldots$, and
$X_{\ell-1}$ at time $\ell$.  These paths are represented by the
lower sloped line in Figure \ref{fig1}.   
Define the groups involving the two sets of paths, $F_0=(X_0,X_1,\ldots,X_\ell)$
and $F_1=(\bone,X_0,\ldots,X_{\ell-1})$.  It can be seen that $F_1\lhd F_0$.
The \csm\ will be seen to be related to the quotient group $F_0/F_1$.
Note that the quotient groups of respective components, e.g., $X_j/X_{j-1}$,
contain a complete set of coset representatives of $X_\ell=B$.

Consider the portion of the trellis given by the following projection:
$$
P_{[0,\ell]}=\chi_{[0,\ell]}(C_{[0,\infty)}).
$$
We call this portion of the trellis, $P_{[0,\ell]}$, the {\it pletty}.
The branches at time epoch $j$ in the pletty, for $0\le j\le\ell$, are $X_j$.  
We call $X_j$ the $j^{\rm th}$ {\it pletty section}.
Thus a pletty consists of the pletty sections $X_0,X_1,\ldots,X_\ell$.

\begin{prop}
Any pletty is a group, called the {\it group pletty}.
\end{prop}

\begin{prf}
The projection map is a homomorphism, so the image is a group \cite{FT}.
\end{prf}

\begin{prop}
Any pletty section is a group, called the {\it group pletty section}.
\end{prop}

\begin{prf}
$X_j$ is a group.
\end{prf}

The elements of $X_j=X_{j-1}(X_j\cap Y_{\ell-j})$ 
form a subgraph of the group trellis section $B$, where each
input state at time $j$ has $|X_0|$ splitting branches and each 
output state at time $j+1$ has $|X_j\cap Y_0|$ merging branches.
The right states of $X_j$, $X_j^+$, are isomorphic to cosets of $X_j\cap Y_0$.
The left states of $X_j$, $X_j^-$, are isomorphic to cosets of $X_0$.
Then $X_j$ is a subdirect product of $X_j^-\times X_j^+$, a subgroup of
the subdirect product $B$.

Of course we know the right states of $X_j$, $X_j^+$, are
connected to the left states of $X_{j+1}$, $X_{j+1}^-$, in the trellis.  Thus
the pletty sections connect to form the pletty.

Consider Figure \ref{fig2} which is a redrawing and refinement of Figure \ref{fig1}.  
We take the point $X_0$ at time $0$ on the upper sloped line 
and pull it vertically so the
sloped line becomes horizontal.  Now the upper horizontal line in Figure \ref{fig2}
represents the paths that split from the identity state at time $0$.  The
groups below $X_0$ at time $0$ are subgroups of $X_0$; the branches in these
subgroups also split from the identity state at time $0$.  We will show these
branches form paths which merge with paths that split from the identity state
at time $1$.

\begin{prop}
\label{prop5}
For any subsets $G,H$ of $B$, we have $(G\cap H)^+=G^+\cap H^+$ 
and $(G\cap H)^-=G^-\cap H^-$.  Also $(GH)^+=G^+H^+$ and $(GH)^-=G^-H^-$.
\end{prop}

\begin{prop}
\label{prop10}
For $j\ge 0$, $k\ge 0$, such that $j+k\le\ell$, we have
\be
\label{prop10a}
(X_j\cap Y_{k})^+=(X_{j+1}\cap Y_{k-1})^-,
\ee
and 
\be
\label{prop10b}
(X_{j-1}(X_j\cap Y_{k}))^+=(X_j(X_{j+1}\cap Y_{k-1}))^-.
\ee
\end{prop}

\begin{prf}
Proof of (\ref{prop10a}).
By definition, we have $X_j^+=X_{j+1}^-$ and $Y_k^+=Y_{k-1}^-$.
Then (\ref{prop10a}) follows from Proposition \ref{prop5}.
The result (\ref{prop10a}) was previously given
in \cite{LM}.

Proof of (\ref{prop10b}).  This results follows from (\ref{prop10a})
and Proposition \ref{prop5}.
\end{prf}

For $k$ such that $0\le k<\ell$, consider $X_{-1}(X_0\cap Y_k)$, 
a subgroup of $X_0$ of branches which split from the identity at time $0$.
From Proposition \ref{prop10}, 
the branches in $X_{-1}(X_0\cap Y_k)$ at time $0$ form paths 
which merge to $X_k$ at time $k+1$.  This is represented by a horizontal
line from time $0$ to time $k+1$ in Figure \ref{fig2}, for 
$k=0,1,2$, and $\ell-1$.

Now we see that Figure \ref{fig2} compares directly with the \csm\ in
(\ref{csm}).  Each set of paths represented by a line in Figure \ref{fig2}
compares directly to a row in (\ref{csm}).  For the subgroups of
branches on a line of Figure \ref{fig2} are the same as the subgroups
in a row of (\ref{csm}).  And as seen from Proposition \ref{prop10},
the right states of any row in column $j$ of (\ref{csm}) are
connected to the left states of the same row in column $j+1$, $0\le j<\ell$,
in the same way as subgroups on a line of Figure \ref{fig2} are
connected.

It can also be seen that column $j$ of (\ref{csm}), $0\le j\le\ell$,
represents the coset decomposition of
the quotient group $X_j/X_{j-1}$ shown in Figure \ref{fig1} and Figure
\ref{fig2}.  The right states of branches in column $j$ are the left
states of branches in column $j+1$, $0\le j<\ell$.  
Thus (\ref{csm}) gives a ``picture'' 
of the groups that occur in the group pletty, as seen in Figure \ref{fig2}.

The rest of this section contains three subsections.  In the first subsection,
we study one pletty section, a column of the \csm\ in (\ref{csm}),
and in the second subsection two pletty sections.  In the last subsection
we study a sequence of pletty sections.  
We will show that a rectangle condition holds for the \csm, in which the 
quotient group formed using paths from $H_0$ to $H_l$ in Figure \ref{fig2}, and paths
from $J_0$ to $J_l$, is isomorphic to the quotient group of individual components,
e.g., $H_0/J_0$ and $H_l/J_l$.

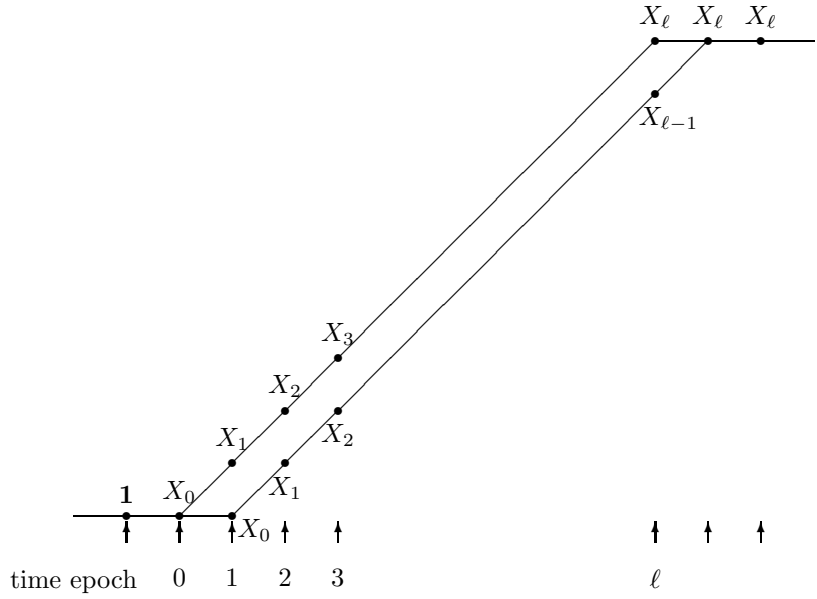
\begin{figure*}[pt]    
\centering
\vspace{3ex}

\begin{picture}(160,200)(-20,-20)

\put(0,0){\line(1,1){180}}
\put(-20,0){\line(1,1){180}}
\put(160,180){\line(1,0){60}}
\put(-60,0){\line(1,0){60}}

\put(-40,0){\circle*{3}}
\put(-20,0){\circle*{3}}
\put(0,0){\circle*{3}}
\put(0,20){\circle*{3}}
\put(20,20){\circle*{3}}
\put(20,40){\circle*{3}}
\put(40,40){\circle*{3}}
\put(40,60){\circle*{3}}
\put(160,160){\circle*{3}}
\put(160,180){\circle*{3}}
\put(180,180){\circle*{3}}
\put(200,180){\circle*{3}}

\put(-40,5){\makebox(0,0)[b]{$\bone$}}
\put(-20,5){\makebox(0,0)[b]{$X_0$}}
\put(2,-5){\makebox(0,0)[l]{$X_0$}}
\put(0,25){\makebox(0,0)[b]{$X_1$}}
\put(20,15){\makebox(0,0)[t]{$X_1$}}
\put(20,45){\makebox(0,0)[b]{$X_2$}}
\put(40,35){\makebox(0,0)[t]{$X_2$}}
\put(40,65){\makebox(0,0)[b]{$X_3$}}
\put(165,155){\makebox(0,0)[t]{$X_{\ell-1}$}}
\put(160,185){\makebox(0,0)[b]{$X_\ell$}}
\put(180,185){\makebox(0,0)[b]{$X_\ell$}}
\put(200,185){\makebox(0,0)[b]{$X_\ell$}}

\put(-40,-10){\vector(0,1){8}}
\put(-20,-10){\vector(0,1){8}}
\put(0,-10){\vector(0,1){8}}
\put(20,-10){\vector(0,1){8}}
\put(40,-10){\vector(0,1){8}}
\put(160,-10){\vector(0,1){8}}
\put(180,-10){\vector(0,1){8}}
\put(200,-10){\vector(0,1){8}}
\put(-60,-20){\makebox(0,0)[t]{\rm{time epoch}}}
\put(-20,-20){\makebox(0,0)[t]{$0$}}
\put(0,-20){\makebox(0,0)[t]{$1$}}
\put(20,-20){\makebox(0,0)[t]{$2$}}
\put(40,-20){\makebox(0,0)[t]{$3$}}
\put(160,-20){\makebox(0,0)[t]{$\ell$}}

\end{picture}

\caption{Trellis paths that split from the identity at time $0$ and time $1$.}
\label{fig1}

\end{figure*}

\begin{figure*}[pb]    
\centering
\vspace{3ex}

\begin{picture}(160,200)(-20,-40)

\put(-20,-20){\line(1,1){200}}
\put(-120,-20){\line(1,0){100}}
\put(-120,0){\line(1,0){120}}
\put(-120,20){\line(1,0){140}}
\put(-120,40){\line(1,0){160}}
\put(-120,90){\line(1,0){210}}
\put(-120,90){\line(1,0){210}}
\put(-120,130){\line(1,0){250}}
\put(-120,160){\line(1,0){280}}
\put(-120,180){\line(1,0){340}}

\put(-20,-20){\circle*{3}}
\put(-120,-20){\circle*{3}}
\put(0,0){\circle*{3}}
\put(-20,0){\circle*{3}}
\put(-120,0){\circle*{3}}
\put(-20,20){\circle*{3}}
\put(-120,20){\circle*{3}}
\put(-20,40){\circle*{3}}
\put(-120,40){\circle*{3}}
\put(-20,90){\circle*{3}}
\put(-120,90){\circle*{3}}
\put(-20,130){\circle*{3}}
\put(-120,130){\circle*{3}}
\put(-20,160){\circle*{3}}
\put(-120,160){\circle*{3}}
\put(0,180){\circle*{3}}
\put(-20,180){\circle*{3}}
\put(-120,180){\circle*{3}}
\put(20,20){\circle*{3}}
\put(20,180){\circle*{3}}
\put(40,40){\circle*{3}}
\put(40,180){\circle*{3}}
\put(70,90){\circle*{3}}
\put(70,130){\circle*{3}}
\put(130,130){\circle*{3}}
\put(90,90){\circle*{3}}
\put(160,160){\circle*{3}}
\put(160,180){\circle*{3}}
\put(180,180){\circle*{3}}
\put(200,180){\circle*{3}}

\put(-15,-17){\makebox(0,0)[br]{$X_{-1}(\bone)$}}
\put(-15,3){\makebox(0,0)[br]{$X_{-1}(X_0\cap Y_0)$}}
\put(-15,23){\makebox(0,0)[br]{$X_{-1}(X_0\cap Y_1)$}}
\put(-15,43){\makebox(0,0)[br]{$X_{-1}(X_0\cap Y_2)$}}
\put(-15,163){\makebox(0,0)[br]{$X_{-1}(X_0\cap Y_{\ell-1})$}}
\put(-15,183){\makebox(0,0)[br]{$X_{-1}(X_0\cap Y_\ell)=X_0$}}

\put(-20,94){\makebox(0,0)[b]{$J_0$}}
\put(-20,134){\makebox(0,0)[b]{$H_0$}}
\put(-18,-23){\makebox(0,0)[l]{$\bone$}}
\put(0,-5){\makebox(0,0)[t]{$X_0$}}
\put(0,184){\makebox(0,0)[b]{$X_1$}}
\put(20,15){\makebox(0,0)[t]{$X_1$}}
\put(20,184){\makebox(0,0)[b]{$X_2$}}
\put(40,35){\makebox(0,0)[t]{$X_2$}}
\put(40,184){\makebox(0,0)[b]{$X_3$}}
\put(70,94){\makebox(0,0)[b]{$J_l$}}
\put(70,134){\makebox(0,0)[b]{$H_l$}}
\put(165,155){\makebox(0,0)[t]{$X_{\ell-1}$}}
\put(160,184){\makebox(0,0)[b]{$X_\ell$}}
\put(180,184){\makebox(0,0)[b]{$X_\ell$}}
\put(200,184){\makebox(0,0)[b]{$X_\ell$}}
\put(-120,-16){\makebox(0,0)[b]{$\bone$}}
\put(-120,4){\makebox(0,0)[b]{$\bone$}}
\put(-120,24){\makebox(0,0)[b]{$\bone$}}
\put(-120,44){\makebox(0,0)[b]{$\bone$}}
\put(-120,94){\makebox(0,0)[b]{$\bone$}}
\put(-120,134){\makebox(0,0)[b]{$\bone$}}
\put(-120,164){\makebox(0,0)[b]{$\bone$}}
\put(-120,184){\makebox(0,0)[b]{$\bone$}}

\put(-120,-30){\vector(0,1){8}}
\put(-20,-30){\vector(0,1){8}}
\put(0,-30){\vector(0,1){8}}
\put(20,-30){\vector(0,1){8}}
\put(40,-30){\vector(0,1){8}}
\put(160,-30){\vector(0,1){8}}
\put(180,-30){\vector(0,1){8}}
\put(200,-30){\vector(0,1){8}}
\put(-60,-40){\makebox(0,0)[t]{\rm{time epoch}}}
\put(-20,-40){\makebox(0,0)[t]{$0$}}
\put(0,-40){\makebox(0,0)[t]{$1$}}
\put(20,-40){\makebox(0,0)[t]{$2$}}
\put(40,-40){\makebox(0,0)[t]{$3$}}
\put(160,-40){\makebox(0,0)[t]{$\ell$}}

\end{picture}

\caption{Rearrangement of Figure \ref{fig1}, showing similarity with
\csm\ (\ref{csm}).}
\label{fig2}

\end{figure*}
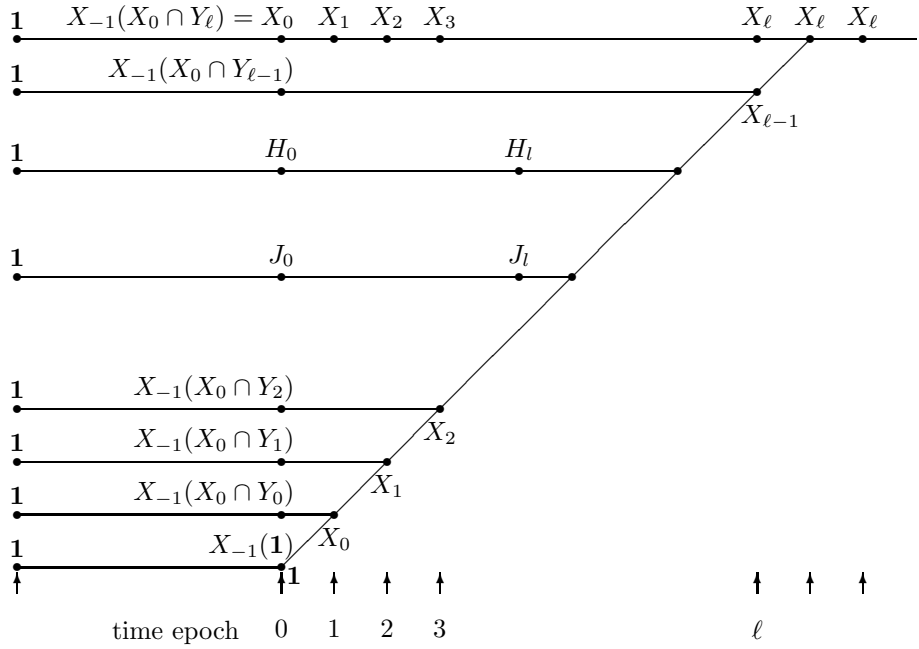

\vspace{3mm}
{\bf 4.1  One pletty section}
\vspace{3mm}

We first state the correspondence theorem since it will be useful
throughout the remainder of Section 4.

\begin{thm}[Correspondence theorem \cite{ROT}]
\label{thm11}
If $K$ is a normal subgroup of $G$ and $\nu:  G\ra G/K$ is the
natural map, then $\nu$ induces a one to one correspondence between
all the subgroups of $G$ that contain $K$ and all the subgroups
of $G/K$.  If $J$ is a subgroup of $G$ that contains $K$, then
$K$ is a normal subgroup of $J$, and the corresponding subgroup of
$G/K$ is $\nu(J)=J/K$.

Let $H$ be another subgroup of $G$ that contains $K$.  Then

(i) $J\le H$ \ifof\ $\nu(J)\le\nu(H)$, and then $[H:J]=[\nu(H):\nu(J)]$;

(ii) $J\lhd H$ \ifof\ $\nu(J)\lhd\nu(H)$, and then $H/J\simeq\nu(H)/\nu(J)$.
\end{thm}

In general, for any groups $Q,Q'$ with $Q'\le Q$, let $Q//Q'$ denote
the right cosets of $Q'$ in $Q$.  For any set $\chi$ and any group $Q$,
we denote a (right) action of $Q$ on $\chi$ by $[\zeta]q$, for $\zeta\in\chi$
and $q\in Q$.

We now study (i) of the correspondence theorem.  
Assume that $J\le H$.  The set of right cosets of $J$ in $H$ 
is denoted by $H//J$.
Any right coset of $H//J$ is of the form $Jh$, for $h\in H$.
The right cosets of $\nu(J)$ in $\nu(H)$ are $\nu(H)//\nu(J)$.  Any
right coset of $\nu(H)//\nu(J)$ is of the form $\nu(Jh)$, where
$Jh$ is a right coset in $H//J$.  Define a function 
$f:  H//J\ra \nu(H)//\nu(J)$ by the assignment 
$Jh\mapsto\nu(Jh)$, for $h\in H$.  Let $g,h\in H$.  
$H$ acts on $H//J$ by the assignment $[Jh]g\mapsto J(hg)$.  
Let $g,h\in H$.  $H$ acts on $\nu(H)//\nu(J)$ by 
the assignment $[\nu(Jh)]g\mapsto \nu(Jhg)$.
 
\begin{cor}
\label{cor12}
We can strengthen (i) of the correspondence theorem this way:  

(i) $J\le H$ \ifof\ $\nu(J)\le\nu(H)$, and then $f$ is an $H$-isomorphism
giving $H//J\simeq \nu(H)//\nu(J)$.
\end{cor}

\begin{prf}
We first show that $f([Jh]g)=[f(Jh)]g$ for $Jh\in H//J$
and $g\in H$.  We have
\begin{multline*}
f([Jh]g)=f(J(hg))=\nu(Jhg) \\
        =[\nu(Jh)]g=[f(Jh)]g.
\end{multline*}
Thus $f$ is an $H$-map.  It is clear that $f$ is a bijection so
$f$ is an $H$-isomorphism.
\end{prf}

We now study a single group pletty section $X_j$, for $0\le j\le\ell$.

\begin{thm}
\label{thm11a}
Consider the $j^{\rm th}$ group pletty section $X_j=X_{j-1}(X_j\cap Y_{\ell-j})$
for $j=0,1,\ldots,\ell$.  Clearly $X_0\subset X_j$ for $j>0$.
For $j=0$, we have $X_0=X_{-1}(X_0\cap Y_\ell)$.  Then $X_0\subset X_j$ for
all $j$, $j=0,1,\ldots,\ell$.  Since $X_0$ is a normal subgroup of $X_j$, 
there is a natural map
$$
\pi_j^-:  X_j\ra X_j/X_0,
$$
and the results of the correspondence theorem apply.  
Note that when $j=0$, $\pi_j^-$ trivially maps to the identity.

For all $j$, $j=0,1,\ldots,\ell$, $X_j\cap Y_0$ is a normal subgroup of $X_j$.
Then there is a natural map
$$
\pi_j^+:  X_j\ra X_j/(X_j\cap Y_0),
$$
and the results of the correspondence theorem apply.
\end{thm}

\begin{thm}
\label{thm17}
Consider the $j^{\rm th}$ group pletty section $X_j=X_{j-1}(X_j\cap Y_{\ell-j})$
for $j=0,1,\ldots,\ell$.  
Fix $j$, $0\le j\le\ell$.
Let $Y'$ and $Y''$ be subgroups of $B$ such that $\bone\le Y'\le Y''\le Y_{\ell-j}$.
Consider the groups $J=X_{j-1}(X_j\cap Y')$ and $H=X_{j-1}(X_j\cap Y'')$,
where $X_{j-1}\le J\le H\le X_j$.

(i) Let 
$$
D=(X_j\cap Y')(X_{j-1}\cap Y'').
$$
$D$ is a subgroup of $X_j\cap Y''$.  There
is a one to one correspondence $\hatf:  H//J\ra (X_j\cap Y'')//D$, and $\hatf$ is
an $H$-isomorphism giving
$$
H//J\simeq (X_j\cap Y'')//D.
$$
We can choose a right transversal of $H//J$ using elements of $X_j\cap Y''$
taken from right cosets of $(X_j\cap Y'')//D$.

(ii) Assume that $Y'\lhd Y''$.  Then $D$ is a normal subgroup of $X_j\cap Y''$.  
There is a one to one correspondence $\hatf:  H/J\ra (X_j\cap Y'')/D$, 
and $\hatf$ is an isomorphism giving
$$
H/J\simeq (X_j\cap Y'')/D.
$$
We can choose a transversal of $H/J$ using elements of $X_j\cap Y''$
taken from cosets of $(X_j\cap Y'')/D$.
\end{thm}

\begin{prf}
$J$ is a group since $X_j\cap Y'$ is a group and $X_{j-1}\lhd B$; similarly
$H$ is a group.  Then $X_{j-1}\le J\le H\le X_j$. 

Proof of (i).
Let $S=X_j\cap Y''$.  Then $H=X_{j-1}(X_j\cap Y'')=JS$.  
We already know $JS$ is a group.  Further $JS$ consists
of all cosets of $J$ that have representatives in $S$.  The 
representatives of $S$ in $J$ are $J\cap S$.  $J\cap S$ is a 
group since $J$ and $S$ are groups, and clearly $J\cap S\subset S$.
The representatives of right coset $Js$ in $S$, where $s\in S$, are 
$(J\cap S)s$ (since $s$ takes $J$ to $Js$ and $J\cap S\subset J$ to
$(J\cap S)s\subset Js$).  Then each right coset $Js$ in $H=JS$ contains
the elements in right coset $(J\cap S)s$ in $S$.  Therefore
there is a one to one correspondence $\hatf:  H//J\ra S//(J\cap S)$, 
where $\hatf$ is an $H$-isomorphism giving
$$
H//J\simeq S//(J\cap S).
$$
Lastly we evaluate $J\cap S$,
\begin{align*}
J\cap S &=X_{j-1}(X_j\cap Y')\cap (X_j\cap Y'') \\
        &=(X_j\cap Y')X_{j-1}\cap (X_j\cap Y''),
\end{align*}
where the last equality follows since $X_{j-1}\lhd B$. 
Let $M=X_j\cap Y'$, $N=X_{j-1}$, and $L=X_j\cap Y''$.
From the Dedekind law, we know $MN\cap L=M(N\cap L)$.
Then 
$$
J\cap S=(X_j\cap Y')(X_{j-1}\cap Y'').
$$
Now define $D$ to be $J\cap S$.

Proof of (ii).
This is an application of the proof of the Zassenhaus lemma
(cf. p. 100 of \cite{ROT}): see Lemma \ref{lem49} below.
\end{prf}

\begin{lem}
\label{lem49}
{\bf (taken from proof of Zassenhaus lemma in \cite{ROT})}
Let $U\lhd U^*$ and $V\lhd V^*$ be four subgroups of a group $G$.
Then $D=(U^*\cap V)(U\cap V^*)$ is a normal subgroup of $U^*\cap V^*$.
If $g\in U(U^*\cap V^*)$, then $g=uu^*$ for $u\in U$ and $u^*\in U^*\cap V^*$.
Define function $f:  U(U^*\cap V^*)\ra (U^*\cap V^*)/D$ by $f(g)=f(uu^*)=Du^*$.
Then $f$ is a well defined homomorphism with kernel $U(U^*\cap V)$ and
$$
\frac{U(U^*\cap V^*)}{U(U^*\cap V)}\simeq\frac{U^*\cap V^*}{D}.
$$
\end{lem}

\vspace{3mm}
{\bf 4.2  Two pletty sections}
\vspace{3mm}

\newcommand{\caln}{\mathcal{N}}
\newcommand{\calp}{\mathcal{P}}

We now study two pletty sections and the Schreier matrix.
We give several analogs of the correspondence theorem for two pletty sections.
We relate adjacent columns in (\ref{csm}).  Fix $j$ such that $0\le j<\ell$.  
Consider the $j^{\rm th}$ group pletty section $X_j$ 
and $(j+1)^{\rm th}$ group pletty section $X_{j+1}$.

For each branch $b\in B$, we define the {\it next branch set} $\caln(b)$
to be the set of branches that can follow $b$ at the next
time epoch in valid trellis paths.  
In other words, branch $e\in\caln(b)$ \ifof\ $b^+=e^-$.

For a set $U\subset B$, define the set $\caln(U)$ to be
the union $\cup_{b\in U} \caln(b)$.  The set $\caln(U)$ always
consists of cosets of $X_0$.  Note that $\caln(X_j)=X_{j+1}$.

\begin{prop}
\label{prop14}
If $b^+=e^-$, the next branch set $\caln(b)$ of 
a branch $b$ in $B$ is the coset $X_0e$ in $B$.
If $b^+=e^-$, the next branch set $\caln(b)$ of a branch $b$ in $X_j$ is the coset
$X_0e$ in $X_{j+1}$.  
\end{prop}

We know $X_j\cap Y_0$ is a normal subgroup of $X_j$, and so there is a
natural map $\pi_j^+:  X_j\ra X_j/X_j\cap Y_0$.
We know $X_0$ is a normal subgroup of $X_{j+1}$, and so there is a natural map
$\pi_{j+1}^-:  X_{j+1}\ra X_{j+1}/X_0$.  

Note that there is an isomorphism
$\alpha_j$,
$$
\alpha_j:  \frac{X_j(X_j\cap Y_0)}{X_j\cap Y_0}\ra\frac{X_jY_0}{Y_0},
$$
or what is the same,
$$
\alpha_j:  \frac{X_j}{X_j\cap Y_0}\ra\frac{X_jY_0}{Y_0},
$$
where $X_jY_0/Y_0$ is considered as a subgroup of $G/Y_0$.
We already know an isomorphism $\varphi$, $\varphi:  G/Y_0\ra G/X_0$,
which gives an isomorphism $\varphi_j$ restricted to $X_jY_0/Y_0$,
$$
\varphi_j:  \frac{X_jY_0}{Y_0}\ra\frac{X_{j+1}}{X_0}.
$$
The composition $\varphi_j\circ\alpha_j$ is an isomorphism, defined to be $\Phi_j$,
$$
\Phi_j:  \frac{X_j}{X_j\cap Y_0}\ra\frac{X_{j+1}}{X_0}.
$$
Then the next branch set $\caln:  X_j\ra X_{j+1}$ 
represents the contraction, isomorphism, and expansion:
$$
X_j{\stackrel{\pi_j^+}{\ra}}X_j/(X_j\cap Y_0)
{\stackrel{\Phi_j}{\simeq}}X_{j+1}/X_0{\stackrel{(\pi_{j+1}^-)^{-1}}{\ra}}X_{j+1}.
$$
Specifically, for branches $b,e$ such that $b^+=e^-$, we have
$\caln(b)=X_0e$, given by the assignments:
$$
b{\stackrel{\pi_j^+}{\mapsto}}(X_j\cap Y_0)b
{\stackrel{\Phi_j}{\mapsto}}X_0e{\stackrel{(\pi_{j+1}^-)^{-1}}{\mapsto}}X_0e.
$$

\begin{prop}
\label{prop40}
There are five results:

(i) Assume $J$ is a set in $X_j$. 
If $J$ is a group, then $\caln(J)$ is a group.  In general,
$\caln(J)$ being a group does not mean set $J$ is a group.
But if set $J$ consists of cosets of $X_j\cap Y_0$, then $J$ 
is a group if $\caln(J)$ is a group.

(ii) Assume $J,H$ are sets in $X_j$.  If $J\le H$, then $\caln(J)\le\caln(H)$.
Assume $J,H$ are groups in $X_j$.  If $J\lhd H$, then $\caln(J)\lhd\caln(H)$.

(iii) Assume $J,H$ are sets in $X_j$ and $\caln(J),\caln(H)$ are sets in $X_{j+1}$.  
In general, $\caln(J)\le\caln(H)$ does not mean $J\le H$.
But if sets $J,H$ consist of cosets of $X_j\cap Y_0$, then $J\le H$
if $\caln(J)\le\caln(H)$.

(iv) Assume $J,H$ are sets in $X_j$ and $\caln(J),\caln(H)$ are groups in $X_{j+1}$.  
If sets $J,H$ consist of cosets of $X_j\cap Y_0$, then $J,H$ are groups. 
And if $\caln(J)\le\caln(H)$, then $J\le H$.
And if $\caln(J)\lhd\caln(H)$, then $J\lhd H$.

(v) Assume $J,H$ are groups in $X_j$.  Assume that $X_j\cap Y_0\subset J$
and $X_j\cap Y_0\subset H$.  Then we
have $J\le H$ \ifof\ $\caln(J)\le\caln(H)$, and 
$J\lhd H$ \ifof\ $\caln(J)\lhd\caln(H)$.
\end{prop}

\begin{prf}
Proof of (i).
First we show that $\caln(J)$ is a group if $J$ is a group.
We know that $\caln(J)=\cup_{g\in J}\caln(g)$.  Let $g'\in\caln(J)$.
Then $g'\in\caln(g)$ for some $g$.  Then $(g,g')$ is a trellis path
segment of length 2.  Similarly let $h'\in\caln(J)$.  Then for some $h$,
there is a trellis path segment $(h,h')$ of length 2.  We know 
$(g,g')*(h,h')=(gh,g'h')$ is a trellis path segment of length 2.
But then $g'h'\in\caln(gh)$, where $gh\in J$.  This means 
$g'h'\in\caln(J)$ and shows $\caln(J)$ is a group.

Now assume $J$ consists of cosets of $X_j\cap Y_0$.  We show
$J$ is a group if $\caln(J)$ is a group.  If $\caln(J)$ is a group,
then the left states $(\caln(J))^-$ are a group.  Since 
$J^+=(\caln(J))^-$, the right states of $J$ are a group.  Then
if $J$ consists of cosets of $X_j\cap Y_0$, $J$ is a group.

Proof of (ii).
Assume $J,H$ are groups and $J\lhd H$.  We show $\caln(J)\lhd\caln(H)$.
Let $h'\in\caln(H)$ and $g'\in\caln(J)$.  We show 
$h'g'(h')^{-1}\in\caln(J)$.  We can find $h\in H$ and $g\in J$ such
that $(h,h')$ and $(g,g')$ are trellis path segments of length 2.
But
$$
(h,h')*(g,g')*(h,h')^{-1}=(hgh^{-1},h'g'(h')^{-1})
$$
is a trellis path segment of length 2.  But $hgh^{-1}\in J$.  Therefore
$h'g'(h')^{-1}\in\caln(J)$.

Proof of (iii).
Assume sets $J,H$ consist of cosets of $X_j\cap Y_0$.  We show $J\le H$
if $\caln(J)\le\caln(H)$.  If $\caln(J)\le\caln(H)$, then 
$(\caln(J))^{-1}\le (\caln(H))^{-1}$.  This means $J^+\le H^+$.
But if sets $J,H$ consist of cosets of $X_j\cap Y_0$, we must have
$J\le H$.

Proof of (iv).
If $\caln(J),\caln(H)$ are groups, then $(\caln(J))^{-1},(\caln(H))^{-1}$
are groups, and $J^+,H^+$ are groups.  But if $J,H$ consist of cosets 
of $X_j\cap Y_0$, this means $J,H$ are groups. 

Assume $\caln(J)\lhd\caln(H)$.  We show $J\lhd H$.  Let $h\in H$ and
$g\in J$.  We show $hgh^{-1}\in J$.  We can find $h'\in\caln(H)$ 
and $g'\in\caln(J)$ such that $(h,h')$ and $(g,g')$ are trellis path 
segments of length 2.  But
$$
(h,h')*(g,g')*(h,h')^{-1}=(hgh^{-1},h'g'(h')^{-1})
$$
is a trellis path segment of length 2.  But $h'g'(h')^{-1}\in\caln(J)$.
If $J$ consists of cosets of $X_j\cap Y_0$, this means $hgh^{-1}\in J$.

Proof of (v).
This follows from (i), (ii), and (iv).
\end{prf}

\begin{prop} 
If $G,H$ are subsets of $B$, then $\caln(GH)=\caln(G)\caln(H)$.
\end{prop}

\begin{prf}
Let $(g,g')$ be a trellis path section of length 2, with $g\in G$,
and similarly for $(h,h')$.  Then $(g,g')*(h,h')=(gh,g'h')$ is a
trellis path section of length 2, with $gh\in GH$.  Then 
$\caln(gh)=\caln(g)\caln(h)$, so $\caln(GH)=\caln(G)\caln(H)$.
\end{prf}

This result means $\caln(Gb)=\caln(G)\caln(b)$ and $\caln(bG)=\caln(b)\caln(G)$ 
for right coset $Gb$ and left coset $bG$.

The following theorem is an analogy of the correspondence theorem,
Theorem \ref{thm11}, and Corollary \ref{cor12}, with the 
mapping $\caln$ in place of the homomorphism $\nu$.

\begin{thm}
\label{thm23}
Fix $j$ such that $0\le j<\ell$.  Let $J$ be a subgroup of $X_j$.  
Then $\caln(J)$ is a group, a subgroup of $X_{j+1}$.  
Let $Jx$ be a right coset of $J$ in $X_j$.  Then $\caln(Jx)$ is a right
coset of $\caln(J)$ in $X_{j+1}$, and all right cosets of $\caln(J)$
in $X_{j+1}$ are of this form.  Define a function 
$q:  X_j//J\ra X_{j+1}//\caln(J)$ by the assignment $q:  Jx\mapsto\caln(Jx)$.
In general, the assignment $q$ gives a many to one correspondence 
between all the right cosets of
$X_j//J$ and all the right cosets of $X_{j+1}//\caln(J)$.
In case $X_j\cap Y_0\subset J$, this becomes a one to one 
correspondence.  In case $J=X_j\cap Y_0$, 
$q$ induces a one to one correspondence between all the subgroups
of $X_j$ that contain $X_j\cap Y_0$ and all the subgroups of $X_{j+1}$ that
contain $X_0$.  If $H$ is a subgroup of $X_j$ that contains $X_j\cap Y_0$,
the corresponding subgroup  of $X_{j+1}//X_0$ is 
$q(H//(X_j\cap Y_0))=\caln(H)//X_0$.

Assume now that $J$ and $H$ are subgroups of $X_j$.

(i) If $J\le H$ then $\caln(J)\le\caln(H)$, and then 
$q|_H:  H//J\ra\caln(H)//\caln(J)$ is an 
$H$-map ($q|_H$ is the mapping $q$ restricted to $H$). 
If in addition $X_j\cap Y_0\subset J$, then $q|_H$ is an $H$-isomorphism.  

(ii) Assume that $X_j\cap Y_0\subset J$ and $X_j\cap Y_0\subset H$.  
We have $J\le H$ \ifof\ $\caln(J)\le\caln(H)$, 
and then $q|_H$ is an $H$-isomorphism.

(iii) If $J\lhd H$ then $\caln(J)\lhd\caln(H)$, and then 
$q|_H:  H/J\ra\caln(H)/\caln(J)$ is a homomorphism.  If in addition
$X_j\cap Y_0\subset J$, then $q|_H$ is an isomorphism.  

(iv) Assume that $X_j\cap Y_0\subset J$ and $X_j\cap Y_0\subset H$.  
We have $J\lhd H$ \ifof\ $\caln(J)\lhd\caln(H)$, 
and then $q|_H$ is an isomorphism.
\end{thm}

\begin{prf}
We know that $\caln(J)=\cup_{g\in J}\caln(g)$.
For each $g\in J$, let $(g,g')$ be a trellis path segment of length 2.
Then $\caln(J)=\cup_{g\in J}\caln(g)=\cup_{g\in J}X_0g'$.

We show that any right coset of $\caln(J)$ in $X_{j+1}$ is of
the form $\caln(Jx)$ where $Jx$ is a right coset of $J$ in $X_j$.  
Let $(b,b')$ be a trellis path segment of length 2 in
$P_{[j,j+1]}$.  Then $(\caln(J))b'$ is a right coset of 
$\caln(J)$.  But we have
\begin{align*}
(\caln(J))b' &=(\cup_{g\in J}X_0g')b' \\
&=\cup_{g\in J}X_0g'b' \\
&=\cup_{g\in J}\caln(gb), \\
&=\caln(Jb),
\end{align*}
where the second from last equality follows since $\caln(gb)=X_0g'b'$.
This result means any right coset of $\caln(J)$ 
is of the form $\caln(Jx)$, where $Jx$ is a right coset of $J$.
This means the function $q$ is well defined.

Proof of (i).  Assume that $J\le H$.  Let $h\in H$.  
Define a function $f:  H//J\ra \caln(H)//\caln(J)$ by 
the assignment $Jh\mapsto\caln(Jh)$.  Let $g,h\in H$.  
$H$ acts on $H//J$ by the assignment $[Jh]g\mapsto J(hg)$.  
Let $g,h\in H$.  $H$ acts on $\caln(H)//\caln(J)$ by 
the assignment $[\caln(Jh)]g\mapsto \caln(Jhg)$.  

We first show that $f([Jh]g)=[f(Jh)]g$ for $Jh\in H//J$
and $g\in H$.  We have
\begin{multline*}
f([Jh]g)=f(J(hg))=\caln(Jhg) \\
        =[\caln(Jh)]g=[f(Jh)]g.
\end{multline*}
Thus $f$ is an $H$-map.  For $h\in H$, note that
$$
q|_H(Jh)=\caln(Jh)=\caln(J)\caln(h)=f(Jh).
$$
Thus $q|_H$ and $f$ are identical, and so $q|_H$ is an $H$-map.  

Assume $J\le H$ and $X_j\cap Y_0\subset J$.  Then the assignment
$f:  Jh\mapsto \caln(Jh)$ is a bijection.
Then $f$ and $q|_H$ are $H$-isomorphisms.  
\end{prf}

\begin{thm}
\label{thm22}
The chief series $\xj$ of an \ellctl\ group trellis $B$ has a refinement which is
a chief series, given by
\begin{multline}
\label{cut1}
\bone=X_{-1}\lhd X_{-1}^*\lhd X_0\lhd X_0^*\lhd X_1\lhd X_1^*\lhd\cdots \\
\cdots\lhd X_{j-1}\lhd X_{j-1}^*\lhd X_j\lhd X_j^*\lhd X_{j+1}\lhd\cdots \\
\cdots\lhd X_{\ell-1}\lhd X_{\ell-1}^*\lhd X_\ell=B,
\end{multline}
where $X_{j-1}^*=X_{j-1}(X_j\cap Y_0)$ for $0\le j\le\ell$, 
and each $X_{j-1}^*\lhd B$.  We know that $X_{\ell-1}^*=X_\ell$ 
by definition of an \ellctl\ \sm.  Fix $j$ such that $0\le j<\ell$.  
Define a function
$$
q':  \frac{X_j}{X_{j-1}^*}\ra\frac{X_{j+1}}{X_j}
$$
by the assignment $q':  X_{j-1}^*h\mapsto\caln(X_{j-1}^*h)$ for $h\in X_j$.
Function $q'$ gives a one to one correspondence 
$X_{j-1}^*h\mapsto\caln(X_{j-1}^*h)$ between all the right cosets
$X_{j-1}^*h$ of $X_j/X_{j-1}^*$ and all the right cosets
$\caln(X_{j-1}^*h)$ of $X_{j+1}/X_j$.  This correspondence gives
an isomorphism
\be
\label{cut2}
\frac{X_j}{X_{j-1}^*}\simeq\frac{\caln(X_j)}{\caln(X_{j-1}^*)}=\frac{X_{j+1}}{X_j},
\ee
where $\caln(X_j)=X_{j+1}$ and $\caln(X_{j-1}^*)=X_j$.
\end{thm}

\begin{prf}
For $0\le j\le\ell$, note that $X_{j-1}\lhd B$ and $X_j\cap Y_0\lhd B$. 
Therefore $X_{j-1}^*=X_{j-1}(X_j\cap Y_0)\lhd B$.

Fix $j$ such that $0\le j<\ell$.  
We apply Theorem \ref{thm23} with $J=X_{j-1}^*$ and $H=X_j$.  Define
a mapping $q':  H/J\ra\caln(H)/\caln(J)$ 
by the assignment $q':  Jh\mapsto\caln(Jh)$.  The quotient group
$\caln(H)/\caln(J)$ is well defined since $\caln(J)\lhd\caln(H)$
if $J\lhd H$.  Since $X_j\cap Y_0\subset X_{j-1}^*=J$, 
from Theorem \ref{thm23} the asssignment $q'$ gives a one to one correspondence 
between all the cosets $X_{j-1}^*h$ of $X_j/X_{j-1}^*$ and all 
cosets $\caln(X_{j-1}^*h)=X_j\caln(h)$ of $X_{j+1}/X_j$.

We have
$$
\caln(J)=\caln(X_{j-1}^*)=\caln(X_{j-1})\caln(X_j\cap Y_0)=X_j.
$$
Then $q'$ is the mapping
$$
q':  \frac{X_j}{X_{j-1}^*}\ra\frac{X_{j+1}}{X_j}.
$$
Since $X_j\cap Y_0\subset X_{j-1}^*=J$, the condition in (iii)
of Theorem \ref{thm23} is met.  Therefore $q'$ is an isomorphism giving
$X_j/X_{j-1}^*\simeq X_{j+1}/X_j$.  This is (\ref{cut2}). 
\end{prf}

\begin{thm}
\label{thm22a}
The \sm\ (\ref{csm}) is a chief series of $B$ which is a refinement of 
the chief series in (\ref{cut1}).  Fix $j$ such that $0\le j<\ell$.
Let $J,H$ be groups such that $X_{j-1}^*\le J\le H\le X_j$.  Define 
function $\psi:  H//J\ra\caln(H)//\caln(J)$
by the assignment $\psi:  Jh\mapsto\caln(Jh)$.  The assignment $\psi$
gives a one to one correspondence between all the right cosets 
$Jh$ of $H//J$ and all the right cosets $\caln(Jh)$ of $\caln(H)//\caln(J)$.
Moreover $\psi$ is an $H$-isomorphism giving 
$$
H//J\simeq\caln(H)//\caln(J).
$$

By (v) of Proposition \ref{prop40},
we have $J\lhd H$ \ifof\ $\caln(J)\lhd\caln(H)$.  Thus if 
$J\lhd H$ or $\caln(J)\lhd\caln(H)$, then $\psi$ is an isomorphism
giving $H/J\simeq\caln(H)/\caln(J)$.
In particular if $H=X_{j-1}(X_j\cap Y_k)$ and $J=X_{j-1}(X_j\cap Y_{k-m})$,
where $k\ge 1$ such that $j+k\le\ell$ and $m\ge 1$ such that $k-m\ge 0$, 
then $\psi$ gives an isomorphism
\begin{align}
\nonumber
\frac{X_{j-1}(X_j\cap Y_k)}{X_{j-1}(X_j\cap Y_{k-m})} &\simeq
\frac{\caln(X_{j-1}(X_j\cap Y_k))}{\caln(X_{j-1}(X_j\cap Y_{k-m}))} \\
\label{new4a}
&=\frac{X_j(X_{j+1}\cap Y_{k-1})}{X_j(X_{j+1}\cap Y_{k-m-1})}.
\end{align}
This is an isomorphism between two adjacent columns of the \sm,
for numerator (denominator) terms in the same row.
\end{thm}

\begin{prf}
The equality in (\ref{new4a}) follows from Proposition \ref{prop10}
and Proposition \ref{prop14}.
\end{prf}

For a set $U\subset B$, define $U\Join\caln(U)$ to be the subset
of $U\times\caln(U)$ consisting of all possible trellis path segments
$(b,b')$ of length 2 that start with branches in $U$.  In other words,
$(b,b')\in U\Join\caln(U)$ \ifof\ $b\in U$, $b'\in\caln(U)$, and
$(b')^-=b^+$.  Define $X_j\Join\caln(X_j)=X_j\Join X_{j+1}$ to be $P_{[j,j+1]}$.

\begin{thm}
\label{thm27}
Fix $j$ such that $0\le j<\ell$.  Let $J$ be a subgroup of $X_j$.  
Then $J\Join\caln(J)$ is a group, a subgroup of $X_j\Join X_{j+1}=P_{[j,j+1]}$.  
Let $Jx$ be a right coset of $J$ in $X_j$.  Then $Jx\Join\caln(Jx)$ is a right
coset of $J\Join\caln(J)$ in $P_{[j,j+1]}$, and all right cosets of 
$J\Join\caln(J)$ in $P_{[j,j+1]}$ are of this form.  Define a function 
$q:  X_j//J\ra P_{[j,j+1]}//(J\Join\caln(J))$ by the assignment 
$q:  Jx\mapsto Jx\Join\caln(Jx)$.
The assignment $q$ gives a one to one correspondence between all the right cosets of
$X_j//J$ and all the right cosets of $P_{[j,j+1]}//(J\Join\caln(J))$.
In case $J=X_j\cap Y_0$, 
$q$ induces a one to one correspondence between all the subgroups
of $X_j$ that contain $X_j\cap Y_0$ and all the subgroups of $P_{[j,j+1]}$ that
contain $(X_j\cap Y_0)\Join X_0$.  
If $H$ is a subgroup of $X_j$ that contains $X_j\cap Y_0$,
the corresponding subgroup of $P_{[j,j+1]}//((X_j\cap Y_0)\Join X_0)$ is   
$q(H//(X_j\cap Y_0))=H\Join\caln(H)//((X_j\cap Y_0)\Join X_0)$.

Assume now that $J$ and $H$ are subgroups of $X_j$.

(i) We have $J\le H$ \ifof\ $J\Join\caln(J)\le H\Join\caln(H)$, and 
then $q|_H:  H//J\ra H\Join\caln(H)//(J\Join\caln(J))$
is an $H$-isomorphism ($q|_H$ is the mapping $q$ restricted to $H$).

(ii) We have $J\lhd H$ \ifof\ $J\Join\caln(J)\lhd H\Join\caln(H)$, 
and then $q|_H:  H/J\ra H\Join\caln(H)/(J\Join\caln(J))$
is an isomorphism.
\end{thm}

\newcommand{\hatr}{{\hat{r}}}
\newcommand{\hatg}{{\hat{g}}}

\begin{prf}
First we show that $J\Join\caln(J)$ is a group.  It is clear that 
$J\Join\caln(J)=\cup_{g\in J}(g\Join\caln(g))$.  Then any element of
$J\Join\caln(J)$ is of the form $(g,g')$ where $g\in J$, and 
$(g,g')$ is a trellis path segment of length 2.  Let $(e,e')$ and
$(s,s')$ be two elements of $J\Join\caln(J)$.  Then 
$(e,e')*(s,s')=(es,e's')$.  But $es\in J$, and $e's'\in\caln(es)$.
Then $(es,e's')\in\cup_{g\in J}(g\Join\caln(g))=J\Join\caln(J)$, and so
$J\Join\caln(J)$ is a group.
    
For each $g\in J$, let $(g,g')$ be a trellis path segment of length 2.
Then $J\Join\caln(J)=\cup_{g\in J}(g\Join\caln(g))=\cup_{g\in J}(g\Join X_0g')$.

We show that any right coset of $J\Join\caln(J)$ in $P_{[j,j+1]}$ is of
the form $Jx\Join\caln(Jx)$ where $Jx$ is a right coset of $J$ in $X_j$.  
Let $(b,b')$ be a trellis path segment of length 2 in
$P_{[j,j+1]}$.  Then $(J\Join\caln(J))*(b,b')$ is a right coset of 
$J\Join\caln(J)$.  But we have
\begin{align*}
(J\Join\caln(J))*(b,b') &=(\cup_{g\in J}(g\Join X_0g'))*(b,b') \\
&=\cup_{g\in J}((g\Join X_0g')*(b,b')) \\
&=\cup_{g\in J}(gb\Join X_0g'b') \\
&=\cup_{g\in J}(gb\Join\caln(gb)),
\end{align*}
where the last equality follows since $\caln(gb)=X_0g'b'$.  We now show
$$
\cup_{g\in J}(gb\Join\caln(gb))=Jb\Join\caln(Jb).
$$
First we show LHS $\subset$ RHS.  Fix $g\in J$.  Then 
$gb\Join\caln(gb)\in Jb\Join\caln(Jb)$.  Now we show RHS $\subset$ LHS.
Any element of $Jb\Join\caln(Jb)$ is of the form $(gb,r')$ for some
$g\in J$ and $r'\in\caln(gb)$.  But then 
$(gb,r')\in gb\Join\caln(gb)\subset \cup_{g\in J}(gb\Join\caln(gb))$.

Combining the above results gives
$$
(J\Join\caln(J))*(b,b')=Jb\Join\caln(Jb).
$$
This result means any right coset of $J\Join\caln(J)$ 
is of the form $Jx\Join\caln(Jx)$, where $Jx$ is a right coset of $J$.
This means the function $q$ is well defined.  Further it is easy to 
see the assignment $Jx\mapsto Jx\Join\caln(Jx)$ 
gives a one to one correspondence between all the right cosets of
$X_j//J$ and all the right cosets of $P_{[j,j+1]}//(J\Join\caln(J))$.

Proof of (i).  Let $J,H$ be subgroups of $X_j$.  Clearly 
$H\Join\caln(H)$ is a group in the same way as $J\Join\caln(J)$ is a group.  
Assume that $J\le H$.  Then clearly   
$J\Join\caln(J)\le H\Join\caln(H)$.  Conversely if
$J\Join\caln(J)\le H\Join\caln(H)$, then we must have $J\le H$.

Assume that $J\le H$.  
Let $h\in H$.  Define a function
$f:  H//J\ra H\Join\caln(H)//(J\Join\caln(J))$ by 
the assignment $Jh\mapsto Jh\Join\caln(Jh)$.  Let $g,h\in H$.  
$H$ acts on $H//J$ by the assignment $[Jh]g\mapsto J(hg)$.  
Let $g,h\in H$.  $H$ acts on $H\Join\caln(H)//(J\Join\caln(J))$ 
by the assignment $[Jh\Join\caln(Jh)]g\mapsto Jhg\Join\caln(Jhg)$.

We show that $f([Jh]g)=[f(Jh)]g$ for $Jh\in H//J$
and $g\in H$:
\begin{multline*}
f([Jh]g)=f(J(hg))=Jhg\Join\caln(Jhg) \\
=[Jh\Join\caln(Jh)]g=[f(Jh)]g.
\end{multline*}
Thus $f$ is an $H$-map.  It is clear from the assignment
$Jh\mapsto Jh\Join\caln(Jh)$ that $f$ is a bijection.
Thus $f$ is an $H$-isomorphism.  For $h\in H$, note that
$$
q|_H(Jh)=Jh\Join\caln(Jh)=f(Jh).
$$
Thus $q|_H$ and $f$ are identical, and so $q|_H$ is an $H$-isomorphism.

Proof of (ii).  Assume $J\lhd H$.  
Let $(g,g')\in J\Join\caln(J)$ and $(h,h')\in H\Join\caln(H)$.  Then
$$
(h,h')*(g,g')*(h,h')^{-1}=(hgh^{-1},h'g'(h')^{-1}).
$$
But $hgh^{-1}\in J$, and therefore $(hgh^{-1},h'g'(h')^{-1})$ 
is an element of $J\Join\caln(J)$.  Then
$J\Join\caln(J)\lhd H\Join\caln(H)$.  Conversely if
$J\Join\caln(J)\lhd H\Join\caln(H)$, it is easy to see $J\lhd H$.
\end{prf}

\begin{lem}
\label{lemx}
Let $J,H$ be groups such that $J\le H\le X_j$.  Define function 
$\psi:  H//J\ra H\Join\caln(H)//(J\Join\caln(J))$ by the assignment 
$\psi:  Jh\mapsto Jh\Join\caln(Jh)$.  Theorem \ref{thm27} shows 
$\psi$ is a one to one correspondence between all the right cosets
$Jh$ of $H//J$ and all the right cosets 
$Jh\Join\caln(Jh)$ of $H\Join\caln(H)//(J\Join\caln(J))$.  The function
$\psi$ induces a mapping $\psi':  H//J\ra \caln(H)//\caln(J)$ 
between all the right cosets $Jh$ of $H//J$ and all the right cosets 
$\caln(Jh)$ of $\caln(H)//\caln(J)$.  From Theorem \ref{thm23},
in general the mapping $\psi'$ is  many to one.  But if 
$X_j\cap Y_0\subset J$, then $\psi'$ is one to one, and in this case,
both $\psi$ and $\psi'$ are one to one.
\end{lem}

We can use Lemma \ref{lemx} in the following way.  
Let $J,H$ be groups such that $J\le H\le X_j$.  We give a procedure
to find right transversals of $H//J$, $\caln(H)//\caln(J)$, and 
$H\Join\caln(H)//(J\Join\caln(J))$.  
Let $Jh$ be a right coset of $J$ in $H$.  Let $b$ be a coset 
representative of $Jh$.  Let $b'$ be any branch in $\caln(b)$.
Then $b'$ is a coset representative of $\caln(Jh)$, a right
coset of $\caln(J)$ in $\caln(H)$.  In addition, $(b,b')$ 
is a trellis path segment of length 2, and 
$(b,b')$ is a coset representative of $Jh\Join\caln(Jh)$, a right
coset of $J\Join\caln(J)$ in $H\Join\caln(H)$.

In general the mapping $\psi'$ in Lemma \ref{lemx} is many to one.
Then there exists coset $Jh^*$ such that $\caln(Jh^*)=\caln(Jh)$.
Let $g$ be a coset 
representative of $Jh^*$.  Let $g'$ be any branch in $\caln(g)$.
Then $g'$ is a coset representative of $\caln(Jh^*)=\caln(Jh)$.
And $(g,g')$ is a coset representative of 
$Jh^*\Join\caln(Jh^*)=Jh^*\Join\caln(Jh)$.  Then this procedure
finds two coset representatives $b',g'$ of $\caln(Jh)$.
Note however that if $X_j\cap Y_0\subset J$, 
the map $\psi'$ is one to one and the procedure obtains just
one coset representative of $\caln(Jh)$.  This gives
the following result.

\begin{thm}
Let $J,H$ be groups such that $J\le H\le X_j$.  
Let $T$ be a right transversal of $H//J$.  For each $b\in T$, pick one
and only one element $b'\in\caln(b)$.  Then 
$\{(b,b') | b\in T\}$ is a collection
of trellis path segments of length 2 that form a right transversal of
$H\Join\caln(H)//J\Join\caln(J)$.  Let $T'=\cup_{b\in T}\{b' | b'\in\caln(b)\}$.  
In general $T'$ contains more than one coset representative from each
coset of $\caln(H)//\caln(J)$.  We can always parse $T'$ so that only
one element is taken from each right coset of $\caln(H)//\caln(J)$, 
such that the
parsed $T'$ is a right transversal of $\caln(H)//\caln(J)$.  If
$X_j\cap Y_0\subset J$, then $T'$ is a right transversal of 
$\caln(H)//\caln(J)$ without parsing.
\end{thm}

{\it Remark:}  Note that we can find transversals in a more
judicious way.  By going backwards in time, first finding a coset
representative $b'$ of $\caln(Jh^*)=\caln(Jh)$, and then coset
representatives $b,g$ of $Jh,Jh^*$, respectively, such that
$(b,b')$ and $(g,b')$ are trellis path segments of length 2, we
can always construct a transversal $T'$ that has only one representative 
in each coset, such that $T'$ is a transversal without parsing.

\vspace{3mm}
{\bf 4.3  A sequence of pletty sections}
\vspace{3mm}

Line (\ref{new4a}) of Theorem \ref{thm22a} shows that if we take 
successive elements in two different rows of (\ref{csm}) and form
quotient groups using elements in the same column (e.g., the four
elements form a rectangle), the quotient groups are isomorphic.
This can be regarded as a correspondence theorem for two group pletty sections.
We now use the preceding results to give a more general version of 
the correspondence theorem for a sequence of pletty sections.

For a set $U\subset B$ and integer $i>0$, define $\caln^i(U)$
to be the $i$-fold composition $\caln^i(U)=\caln\circ\caln\circ\cdots\circ\caln(U)$.
For $i=0$, define $\caln^i(U)=\caln^0(U)$ to be just $U$.

For a set $U\subset B$, define $U\Join\caln(U)\Join\caln^2(U)$ to be
$(U\Join\caln(U))\Join\caln^2(U)$, the subset
of $(U\Join\caln(U))\times\caln^2(U)$ consisting of all possible trellis path segments
of length 3 that start with trellis path segments
of length 2 in $U\Join\caln(U)$.  This is just the set
of all possible trellis path segments of length 3 that start 
with branches in $U$.

For a set $U\subset B$, in a similar way 
define $U\Join\caln(U)\Join\cdots\Join\caln^i(U)$.  This is 
just the subset of $U\times\caln(U)\times\cdots\times\caln^i(U)$
consisting of all possible trellis path segments
of length $i+1$ that start with branches in $U$.
             
The next result follows from Proposition \ref{prop10}.

\begin{prop}
\label{prop31}
Fix integer $k$, $0\le k\le\ell$.  For $0\le j\le k$, we have
\be
\label{eqnn1}
(X_j\cap Y_{k-j})^+=(X_{j+1}\cap Y_{k-j-1})^-.
\ee
\end{prop}

\noindent This result means the sequence of groups 
\begin{multline*}
\ldots,\bone,X_0\cap Y_k,\ldots,X_j\cap Y_{k-j},X_{j+1}\cap Y_{k-j-1},  \\
\ldots,X_k\cap Y_0,\bone,\ldots
\end{multline*}
consists of paths which split from the identity state at time epoch $0$ and 
merge to the identity state at time epoch $k+1$.

We choose $H_0$ such that $H_0\le X_{-1}(X_0\cap Y_\ell)$ and $J_0$ such that
$J_0\ge X_{-1}(X_0\cap Y_0)$.  Thus fix integer $k$, $0<k\le\ell$.  
Choose integer $m\ge 1$ such that $k-m>-1$.  Define 
\begin{align*}
H_0 &=X_{-1}(X_0\cap Y_k), \\
J_0 &=X_{-1}(X_0\cap Y_{k-m}).
\end{align*}
Choose integer $l$ such that $l>0$ and $l\le k-m+1$.
For $0<j\le l$, recursively define
\begin{align*}
H_j &=\caln(H_{j-1}), \\
J_j &=\caln(J_{j-1}).
\end{align*}
Then using Proposition \ref{prop31} and Proposition \ref{prop14},
\begin{align*}
H_j &=X_{j-1}(X_j\cap Y_{k-j}), \\
J_j &=X_{j-1}(X_j\cap Y_{k-j-m}).
\end{align*}
For $0\le j\le l$, the groups $H_j$ and $J_j$ correspond to rows in (\ref{csm})
or horizontal lines in Figure \ref{fig2}.  Note that for $l=k-m+1$,
$J_l=X_{l-1}(\bone)=X_{k-m}(\bone)$ is the group on the diagonal of (\ref{csm})
or the diagonal of Figure \ref{fig2}.

For $0\le j\le l$, $H_j$ and $J_j$ are subgroups of $X_j$. 
For $0<j\le l$, $H_j$ and $J_j$ contain $X_0$.  For $0\le j\le l$,
$H_j$ contains $X_j\cap Y_0$.  For $l<k-m+1$ and $0\le j\le l$,
$J_j$ contains $X_j\cap Y_0$.  For $l=k-m+1$ and $0\le j<l$,
$J_j$ contains $X_j\cap Y_0$.

Define
$$
P_{[0,l]}=X_0\Join\caln(X_0)\Join\caln^2(X_0)\Join\cdots\Join\caln^l(X_0),
$$
$$
H_{[0,l]}=H_0\Join\caln(H_0)\Join\caln^2(H_0)\Join\cdots\Join\caln^l(H_0),
$$
and
$$
J_{[0,l]}=J_0\Join\caln(J_0)\Join\caln^2(J_0)\Join\cdots\Join\caln^l(J_0).
$$
These are just
$$
P_{[0,l]}=X_0\Join X_1\Join X_2\Join\cdots\Join X_l,
$$
$$
H_{[0,l]}=H_0\Join H_1\Join H_2\Join\cdots\Join H_l,
$$
and
$$
J_{[0,l]}=J_0\Join J_1\Join J_2\Join\cdots\Join J_l.
$$
$P_{[0,l]}$ is a sequence of pletty sections $X_0,X_1,\ldots,X_l$.

In Theorem \ref{thma}, we give an extension of the correspondence theorem 
to a \csm.  This result can be regarded as a rectangle criterion for a \csm, 
with $H_0$, $J_0$, $H_l$, and $J_l$ as the corners of a rectangle, as shown in
Figure \ref{fig2}.  It is similar in spirit to a quadrangle criterion for
a Latin square \cite{DK} or a configuration theorem for a net \cite{DJ}.

\begin{thm}
\label{thma}
We have $P_{[0,l]}$, $H_{[0,l]}$, and $J_{[0,l]}$ are groups with
$J_{[0,l]}<H_{[0,l]}\le P_{[0,l]}$.

Let $J_0h_0$ be a right coset of $J_0$ in $H_0$.  Then
$$
J_0h_0\Join\caln(J_0h_0)\Join\caln^2(J_0h_0)\Join\cdots\Join\caln^l(J_0h_0)
$$
is a right coset of $J_{[0,l]}$ in $H_{[0,l]}$.  And for $j$ such that
$0\le j\le l$, $\caln^j(J_0h_0)$ is a right coset of $J_j$ in $H_j$.  
There are 4 results:

(i) The function $f: H_0//J_0\ra H_{[0,l]}//J_{[0,l]}$ defined by the 
assignment 
$$
f:  J_0h_0\mapsto J_0h_0\Join\caln(J_0h_0)\Join\caln^2(J_0h_0)\Join\cdots\Join\caln^l(J_0h_0)
$$
gives a one to one correspondence between all the right cosets of 
$H_0//J_0$ and all the right cosets of $H_{[0,l]}//J_{[0,l]}$.

We have $J_0\le H_0$ \ifof\ $J_{[0,l]}\le H_{[0,l]}$,  
and then $f$ is an $H_0$-isomorphism giving
$$
H_0//J_0\simeq H_{[0,l]}//J_{[0,l]}.
$$

We have $J_0\lhd H_0$ \ifof\ $J_{[0,l]}\lhd H_{[0,l]}$,  
and then $f$ is an isomorphism giving
$$
H_0/J_0\simeq H_{[0,l]}/J_{[0,l]}.
$$

(ii) Provided $l>0$, fix $j$ such that $0<j\le l$. The function 
$f_{0,j}: H_0//J_0\ra H_j//J_j$ defined by the assignment
$$
f_{0,j}:  J_0h_0\mapsto\caln^j(J_0h_0)
$$ 
gives a one to one correspondence between all the right cosets of 
$H_0//J_0$ and all the right cosets of $H_j//J_j$.

We have $J_0\le H_0$ \ifof\ $J_j\le H_j$,
and then $f_{0,j}$ is an $H_0$-isomorphism giving
$$
H_0//J_0\simeq H_j//J_j.
$$

We have $J_0\lhd H_0$ \ifof\ $J_j\lhd H_j$,
and then $f_{0,j}$ is an isomorphism giving
$$
H_0/J_0\simeq H_j/J_j.
$$

(iii) Provided $l>0$, fix $j,k$ such that $0\le j<k\le l$. 
The function $f_{j,k}: H_j//J_j\ra H_k//J_k$ defined by the assignment
$$
f_{j,k}:  \caln^j(J_0h_0)\mapsto\caln^k(J_0h_0)
$$ 
gives a one to one correspondence between all the right cosets of 
$H_j//J_j$ and all the right cosets of $H_k//J_k$.

We have $J_j\le H_j$ \ifof\ $J_k\le H_k$,  
and then $f_{j,k}$ is an $H_j$-isomorphism giving
$$
H_j//J_j\simeq H_k//J_k.
$$

We have $J_j\lhd H_j$ \ifof\ $J_k\lhd H_k$,  
and then $f_{j,k}$ is an isomorphism giving
$$
H_j/J_j\simeq H_k/J_k.
$$

(iv) Fix $j$ such that $0\le j\le l$. 
The function $f_j^*: H_{[0,l]}//J_{[0,l]}\ra H_j//J_j$
defined by the assignment
\begin{align*}
f_j^*:  J_0h_0\Join\caln(J_0h_0)\Join\caln^2(J_0h_0)\Join\cdots\Join&\caln^l(J_0h_0) \\
&\mapsto\caln^j(J_0h_0)
\end{align*}
gives a one to one correspondence between all the right cosets of 
$H_{[0,l]}//J_{[0,l]}$ and all the right cosets of $H_j//J_j$.

We have $J_j\le H_j$ \ifof\ $J_{[0,l]}\le H_{[0,l]}$,
and then $f_j^*$ is an $H_{[0,l]}$-isomorphism giving
$$
H_{[0,l]}//J_{[0,l]}\simeq H_j//J_j.
$$

We have $J_j\lhd H_j$ \ifof\ $J_{[0,l]}\lhd H_{[0,l]}$,
and then $f_j^*$ is an isomorphism giving
$$
H_{[0,l]}/J_{[0,l]}\simeq H_j/J_j.
$$
\end{thm}

\begin{prf}
First we show that $J_0\Join\caln(J_0)\Join\cdots\Join\caln^l(J_0)$
is a group.  It is clear that 
\begin{align*}
J_0\Join\caln(J_0)&\Join\cdots\Join\caln^l(J_0) \\
&=\cup_{g\in J_0}(g\Join\caln(g)\Join\cdots\Join\caln^l(g)).
\end{align*}
Then any element of $J_0\Join\caln(J_0)\Join\cdots\Join\caln^l(J_0)$
is of the form $(g,g',\ldots,g'')$ where $g\in J_0$, and $(g,g',\ldots,g'')$ 
is a trellis path segment of length $l+1$.  Let $(e,e',\ldots,e'')$ and
$(s,s',\ldots,s'')$ be two elements of $J_0\Join\caln(J_0)\Join\cdots\Join\caln^l(J_0)$.  Then 
$(e,e',\ldots,e'')*(s,s',\ldots,s'')=(es,e's',\ldots,e''s'')$.  But $es\in J_0$, $e's'\in\caln(es)$,
and $e''s''\in\caln^l(es)$.  Then 
$(es,e's',\ldots,e''s'')\in\cup_{g\in J_0}(g\Join\caln(g)\Join\cdots\Join\caln^l(g))$, and so
$J_0\Join\caln(J_0)\Join\cdots\Join\caln^l(J_0)$ is a group.

For each $g\in J_0$, let $(g,g',\ldots,g'')$ be a trellis path segment of length $l+1$.
Then 
\begin{align*}
J_0\Join\caln(J_0)&\Join\cdots\Join\caln^l(J_0) \\
&=\cup_{g\in J_0}(g\Join\caln(g)\Join\cdots\Join\caln^l(g)) \\
&=\cup_{g\in J_0}(g\Join X_0g'\Join\cdots\Join X_{l-1}g'').
\end{align*}

We show that any right coset of $J_0\Join\caln(J_0)\Join\cdots\Join\caln^l(J_0)$ in $P_{[0,l]}$
is of the form $J_0x\Join\caln(J_0x)\Join\cdots\Join\caln^l(J_0x)$ where $J_0x$ is a right
coset of $J_0$ in $X_0$.
Let $(b,b',\ldots,b'')$ be a trellis path segment of length $l+1$ in
$P_{[0,l]}$.  Then $(J_0\Join\caln(J_0)\Join\cdots\Join\caln^l(J_0))*(b,b',\ldots,b'')$ is a 
right coset of $J_0\Join\caln(J_0)\Join\cdots\Join\caln^l(J_0)$.  But we have
\begin{align*}
(J_0&\Join\caln(J_0)\Join\cdots\Join\caln^l(J_0))*(b,b',\ldots,b'') \\ 
&=(\cup_{g\in J_0}(g\Join X_0g'\Join\cdots\Join X_{l-1}g''))*(b,b',\ldots,b'') \\
&=\cup_{g\in J_0}((g\Join X_0g'\Join\cdots\Join X_{l-1}g'')*(b,b',\ldots,b'')) \\
&=\cup_{g\in J_0}(gb\Join X_0g'b'\Join\cdots\Join X_{l-1}g''b'') \\
&=\cup_{g\in J_0}(gb\Join\caln(gb)\Join\cdots\Join\caln^l(gb)),
\end{align*}
where the last equality follows since $\caln(gb)=X_0g'b'$ and 
$\caln^l(gb)=X_{l-1}g''b''$.  

We now show
\begin{align*}
\cup_{g\in J_0}(gb\Join\caln(gb)&\Join\cdots\Join\caln^l(gb)) \\
&=J_0b\Join\caln(J_0b)\Join\cdots\Join\caln^l(J_0b).
\end{align*}
First we show LHS $\subset$ RHS.  Fix $g\in J_0$.  Then 
$gb\Join\caln(gb)\Join\cdots\Join\caln^l(gb)\in J_0b\Join\caln(J_0b)\Join\cdots\Join\caln^l(J_0b)$.  
Now we show RHS $\subset$ LHS.
Any element of $J_0b\Join\caln(J_0b)\Join\cdots\Join\caln^l(J_0b)$ is of the form $(gb,r',\ldots,r'')$ for some
$g\in J_0$, $r'\in\caln(gb)$, and $r''\in\caln^l(gb)$.  But then 
$(gb,r',\ldots,r'')\in gb\Join\caln(gb)\Join\cdots\Join\caln^l(gb)
\subset \cup_{g\in J_0}(gb\Join\caln(gb)\Join\cdots\Join\caln^l(gb))$.

Combining the above results gives
\begin{align*}
(J_0\Join\caln(J_0)&\Join\cdots\Join\caln^l(J_0))*(b,b',\ldots,b'') \\
&=J_0b\Join\caln(J_0b)\Join\cdots\Join\caln^l(J_0b).
\end{align*}
This result means any right coset of $J_0\Join\caln(J_0)\Join\cdots\Join\caln^l(J_0)$ 
is of the form $J_0x\Join\caln(J_0x)\Join\cdots\Join\caln^l(J_0x)$, 
where $J_0x$ is a right coset of $J_0$.
This means the function $f$ is well defined.  Further it is easy to 
see the function $f$ 
gives a one to one correspondence between all the right cosets of 
$H_0//J_0$ and all the right cosets of $H_{[0,l]}//J_{[0,l]}$.
This proves the first part of (i).  The proof of the remainder of (i)
is similar to the proof of (i) and (ii) of Theorem \ref{thm27}.

Proof of (ii).
We proceed by induction on $n$, $n=1,\ldots,l$.  Consider the following
hypothesis (*):

(*) The function $f_{0,n}: H_0//J_0\ra H_n//J_n$ defined by the assignment
$$
f_{0,n}:  J_0h_0\mapsto\caln^n(J_0h_0)
$$ 
gives a one to one correspondence between all the right cosets of 
$H_0//J_0$ and all the right cosets of $H_n//J_n$.
We have $J_0\le H_0$ \ifof\ $J_n\le H_n$,
and then $f_{0,n}$ is an $H_0$-isomorphism giving
$$
H_0//J_0\simeq H_n//J_n.
$$
We have $J_0\lhd H_0$ \ifof\ $J_n\lhd H_n$,
and then $f_{0,n}$ is an isomorphism giving
$$
H_0/J_0\simeq H_n/J_n.
$$

We know that hypothesis (*) is true for $n=1$, by Theorem \ref{thm23}.
Thus assume hypothesis (*) holds for $n$, $1\le n<l$; we show it holds for $n+1$.

Define a function $q:  H_n//J_n\ra \caln(H_n)//\caln(J_n)$ by 
the assignment $q:  J_nx\mapsto\caln(J_nx)$, for $x\in H_n$.
By Theorem \ref{thm23}, $q$ is a one to one correspondence 
between all right cosets $J_nx$ of $H_n//J_n$ and all right
cosets of $\caln(H_n)//\caln(J_n)=H_{n+1}//J_{n+1}$.  By hypothesis
(*) all right cosets of $H_n//J_n$ are of the form $\caln^n(J_0h_0)$,
$h_0\in H_0$.  Thus there is a one to one correspondence between all
right cosets written as $J_nx$ and all right cosets written as
$\caln^n(J_0h_0)$.  Then $q$ gives the assignment
$$
q:  \caln^n(J_0h_0)\mapsto\caln(\caln^n(J_0h_0))=\caln^{n+1}(J_0h_0).
$$
Thus all right cosets of $H_{n+1}//J_{n+1}$ are of the form
$\caln^{n+1}(J_0h_0)$.
Define $f_{0,n+1}:  H_0//J_0\ra H_{n+1}//J_{n+1}$ by $q\circ f_{0,n}$.
Then $f_{0,n+1}$ gives the assignment 
$f_{0,n+1}:  J_0h_0\mapsto\caln^{n+1}(J_0h_0)$, which is a one to one
correspondence between all the right cosets of $H_0//J_0$ and all
right cosets of $H_{n+1}//J_{n+1}$.

By hypothesis we have $J_0\le H_0$ \ifof\ $J_n\le H_n$, 
and then $f_{0,n}$ is an $H_0$-isomorphism.  From
Theorem \ref{thm23}, since $X_j\cap Y_0\subset J_n$, we have 
$J_n\le H_n$ \ifof\ $J_{n+1}\le H_{n+1}$, and $q$ is an $H_n$-isomorphism.
Then $J_0\le H_0$ \ifof\ $J_{n+1}\le H_{n+1}$.  Since
$f_{0,n}$ is an $H_0$-isomorphism, and $q$ is an $H_n$-isomorphism,
it can be shown $f_{0,n+1}=q\circ f_{0,n}$ is an $H_0$-isomorphism.

The remainder of the proof, that $J_0\lhd H_0$ \ifof\ $J_{n+1}\lhd H_{n+1}$,
and $f_{0,n+1}$ is an isomorphism, is similar.

Proof of (iii).
The function $f_{0,j}$ in (ii) is a bijection and so the inverse
$f_{0,j}^{-1}: H_j//J_j\ra H_0//J_0$ exists.  It is easy to show
$f_{0,j}^{-1}$ is an $H_j$-isomorphism.  Define the function
$f_{j,k}: H_j//J_j\ra H_k//J_k$ by $f_{j,k}=f_{0,k}\circ f_{0,j}^{-1}$.
Then $f_{j,k}$ gives a one to one correspondence between all the right cosets of 
$H_j//J_j$ and all the right cosets of $H_k//J_k$, and 
$f_{j,k}$ is an $H_j$-isomorphism.

Proof of (iv).
The function $f$ in (i) is a bijection and so the inverse
$f^{-1}: H_{[0,l]}//J_{[0,l]}\ra H_0//J_0$ exists, and 
$f^{-1}$ is an $H_{[0,l]}$-isomorphism.  Define the function
$f_j^*$ by $f_j^*=f_{0,j}\circ f^{-1}$.
Then $f_j^*$ gives a one to one correspondence between right cosets
in $H_{[0,l]}//J_{[0,l]}$ and $H_j//J_j$, and 
$f_j^*$ is an $H_{[0,l]}$-isomorphism.
\end{prf}

{\it Remark:}  The theorem holds for a more general 
rectangle, using $H_{l'}$ in place of $H_0$ and $J_{l'}$ in place of $J_0$,
for $0<l'<l$.  This more general result is not needed.

We can diagram Theorem \ref{thma} as shown in Figure \ref{fig3}.
This diagram shows some formal similarity with the trellis analog
\cite{LM} of the code granule theorem \cite{FT} (cf. Theorem 8.1 of \cite{LM}).

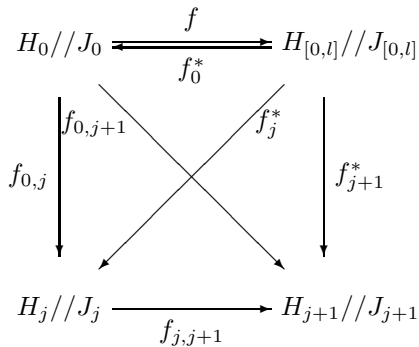
\begin{figure}[h]
\centering
\vspace{3ex}

\begin{picture}(100,100)(0,-5)

\put(0,0){\makebox(0,0){$H_j//J_j$}}
\put(110,0){\makebox(0,0){$H_{j+1}//J_{j+1}$}}
\put(110,100){\makebox(0,0){$H_{[0,l]}//J_{[0,l]}$}}
\put(0,100){\makebox(0,0){$H_0//J_0$}}

\put(20,0){\vector(1,0){60}}
\put(20,101){\vector(1,0){60}}
\put(80,99){\vector(-1,0){60}}
\put(0,80){\vector(0,-1){60}}
\put(15,85){\vector(1,-1){70}}
\put(85,85){\vector(-1,-1){70}}
\put(100,80){\vector(0,-1){60}}

\put(50,-4){\makebox(0,0)[t]{$f_{j,j+1}$}}
\put(50,105){\makebox(0,0)[b]{$f$}}
\put(-4,50){\makebox(0,0)[r]{$f_{0,j}$}}
\put(26,70){\makebox(0,0)[r]{$f_{0,j+1}$}}
\put(50,95){\makebox(0,0)[t]{$f_0^*$}}
\put(74,70){\makebox(0,0)[l]{$f_j^*$}}
\put(104,50){\makebox(0,0)[l]{$f_{j+1}^*$}}

\end{picture}

\caption{Correspondence theorem for $P_{[0,l]}$.  All functions
are bijections.  Functions $f$, $f_{0,j}$, and $f_{0,j+1}$ are $H_0$-isomorphisms.
Functions $f_0^*$, $f_j^*$, and $f_{j+1}^*$ are $H_{[0,l]}$-isomorphisms.  Function
$f_{j,j+1}$ is an $H_j$-isomorphism.  If $J_0\lhd H_0$,
then all sets of right cosets become quotient groups and all bijections
become isomorphisms.}
\label{fig3}

\end{figure}

A coset representative of $H_{[0,l]}//J_{[0,l]}$ is a trellis path
of length $l+1$.  Then a right transversal of $H_{[0,l]}//J_{[0,l]}$ must consist
of trellis paths of length $l+1$.  For $j=0,1,\ldots,l$, by 
{\it component} $j$ of a transversal, we mean the set of branches
which are the $j^{\rm th}$ branch in each of the trellis paths of length
$l+1$.  Note that if we arbitrarily select right transversals of
$H_j//J_j$, for $j=0,1,\ldots,l$, we do not necessarily obtain a
right transversal of $H_{[0,l]}//J_{[0,l]}$, or even trellis paths of
length $l+1$.  However we now show how to construct 
a right transversal of $H_{[0,l]}//J_{[0,l]}$ such that
component $j$ forms a right transversal of $H_j//J_j$, for $j=0,1,\ldots,l$.

We apply Figure \ref{fig3} repeatedly.  First find a transversal
$T_0$ of $H_0//J_0$, $T_0=\{b_0 | b_0\in T_0\}$.
For each $b_0\in T_0$, choose a $b_1\in\caln(b_0)$.
Since the map $f_{0,1}:  H_0//J_0\ra H_1//J_1$ is an
$H_0$-isomorphism, the collection 
$T_1=\cup_{b_0\in T_0}\{b_1 | b_1\in\caln(b_0)\}$ is a right transversal
of $H_1//J_1$, and $\{(b_0,b_1) | b_0\in T_0, b_1\in\caln(b_0)\}$ is a collection
of trellis path segments of length 2 that form a right transversal of
$H_{[0,1]}//J_{[0,1]}$.   

Next, for each $b_1\in T_1$, choose a $b_2\in\caln(b_1)$.
Since the map $f_{1,2}:  H_1//J_1\ra H_2//J_2$ is an
$H_1$-isomorphism, the collection 
$T_2=\cup_{b_1\in T_1}\{b_2 | b_2\in\caln(b_1)\}$ is a right transversal
of $H_2//J_2$, and 
$\{(b_0,b_1,b_2) | b_0\in T_0, b_1\in\caln(b_0), b_2\in\caln(b_1)\}$ 
is a collection of trellis path segments of length 3 that form a 
right transversal of $H_{[0,2]}//J_{[0,2]}$.  Continuing in this
manner gives the following result.

\begin{thm}
\label{thmd}
Fix $l$ such that $0<l\le k-m+1$.  There are two results:

(i) We can select a right transversal of $H_{[0,l]}//J_{[0,l]}$ such that
for $j=0,1,\ldots,l$, component $j$ forms a right transversal of $H_j//J_j$.

(ii) Assume that $J_0\lhd H_0$.  Then $J_i\lhd H_i$, $i>0$, and
$J_{[0,l]}\lhd H_{[0,l]}$.
We can select a transversal of $H_{[0,l]}/J_{[0,l]}$ such that
for $j=0,1,\ldots,l$, component $j$ forms a transversal of $H_j/J_j$.
\end{thm}

We now give an extension of Theorem \ref{thmd}.
First we give a useful extension of Figure \ref{fig3}.
Fix $j$, $0\le j<l$.  From Theorem \ref{thm17}, 
we know there is a one to one correspondence
$\hatf_j:  H_j//J_j\ra (X_j\cap Y_{l-j})//D_j$, where
$$
D_j=(X_j\cap Y_{l-j-m})(X_{j-1}\cap Y_{l-j}).
$$
Also $\hatf_j$ is an $H_j$-isomorphism.  Under the one to one
correspondence, the elements of a right coset of $(X_j\cap Y_{l-j})//D_j$
are contained in a right coset of $H_j//J_j$.
Similarly there is a one to one correspondence
$\hatf_{j+1}:  H_{j+1}//J_{j+1}\ra (X_{j+1}\cap Y_{l-j-1})//D_{j+1}$, where
$$
D_{j+1}=(X_{j+1}\cap Y_{l-j-m-1})(X_j\cap Y_{l-j-1}).
$$
Also $\hatf_{j+1}$ is an $H_{j+1}$-isomorphism.  Under the one to one
correspondence, the elements of a right coset of 
$(X_{j+1}\cap Y_{l-j-1})//D_{j+1}$
are contained in a right coset of $H_{j+1}//J_{j+1}$.
These relations are shown in Figure \ref{fig4}.

\begin{figure}[h]
\centering
\vspace{3ex}

\begin{picture}(100,200)(0,-5)

\put(0,100){\makebox(0,0){$H_j//J_j$}}
\put(110,100){\makebox(0,0){$H_{j+1}//J_{j+1}$}}
\put(110,200){\makebox(0,0){$H_{[0,l]}//J_{[0,l]}$}}
\put(0,200){\makebox(0,0){$H_0//J_0$}}
\put(-20,0){\makebox(0,0){$(X_j\cap Y_{l-j})//D_j$}}
\put(120,0){\makebox(0,0){$(X_{j+1}\cap Y_{l-j-1})//D_{j+1}$}}

\put(20,100){\vector(1,0){60}}
\put(20,201){\vector(1,0){60}}
\put(80,199){\vector(-1,0){60}}
\put(0,180){\vector(0,-1){60}}
\put(15,185){\vector(1,-1){70}}
\put(85,185){\vector(-1,-1){70}}
\put(100,180){\vector(0,-1){60}}
\put(0,80){\vector(0,-1){60}}
\put(20,0){\vector(1,0){45}}
\put(100,80){\vector(0,-1){60}}

\put(50,96){\makebox(0,0)[t]{$f_{j,j+1}$}}
\put(50,205){\makebox(0,0)[b]{$f$}}
\put(-4,150){\makebox(0,0)[r]{$f_{0,j}$}}
\put(26,170){\makebox(0,0)[r]{$f_{0,j+1}$}}
\put(50,195){\makebox(0,0)[t]{$f_0^*$}}
\put(74,170){\makebox(0,0)[l]{$f_j^*$}}
\put(104,150){\makebox(0,0)[l]{$f_{j+1}^*$}}
\put(-4,50){\makebox(0,0)[r]{$\hatf_j$}}
\put(50,-4){\makebox(0,0)[t]{$f_{j,j+1}'$}}
\put(104,50){\makebox(0,0)[l]{$\hatf_{j+1}$}}

\end{picture}

\caption{Properties of $P_{[0,l]}$.  All functions
are bijections.  Functions $f$, $f_{0,j}$, and $f_{0,j+1}$ are $H_0$-isomorphisms.
Functions $f_0^*$, $f_j^*$, and $f_{j+1}^*$ are $H_{[0,l]}$-isomorphisms.  Functions
$f_{j,j+1}$ and $\hatf_j$ are $H_j$-isomorphisms.  Function $\hatf_{j+1}$
is an $H_{j+1}$-isomorphism.  Function $f_{j,j+1}'$ is an 
$(X_j\cap Y_{l-j})$-isomorphism.  If $J_0\lhd H_0$,
then all sets of right cosets become quotient groups and all bijections
become isomorphisms.}
\label{fig4}

\end{figure}
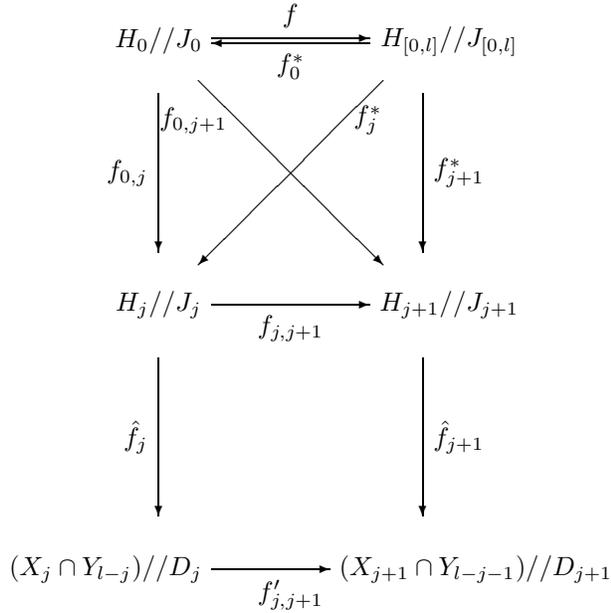

Now note further that 
\be
\label{star}
(X_j\cap Y_{l-j})^+=(X_{j+1}\cap Y_{l-j-1})^-
\ee
and
\be
\label{star1}
D_j^+=D_{j+1}^-.
\ee
Then $H_j$-isomorphism $f_{j,j+1}$ induces a one to one correspondence
$$
f_{j,j+1}':  (X_j\cap Y_{l-j})//D_j\ra (X_{j+1}\cap Y_{l-j-1})//D_{j+1},
$$
and $f_{j,j+1}'$ is an $(X_j\cap Y_{l-j})$-isomorphism (see Figure \ref{fig4}).

Assume we have chosen a right transversal $T$ of $H_j//J_j$ using elements of
$X_j\cap Y_{l-j}$ taken from right cosets of $(X_j\cap Y_{l-j})//D_j$;
we can use function $\hatf_j$ to do this.
For each $b\in T$, choose a $b'\in\caln(b)$ such that 
$b'\in X_{j+1}\cap Y_{l-j-1}$; by (\ref{star}) this
is always possible.  Under the one to one correspondence $f_{j,j+1}'$,
we know the collection $T'=\cup_{b\in T}\{b' | b'\in\caln(b)\}$ is a 
right transversal of $(X_{j+1}\cap Y_{l-j-1})//D_{j+1}$.  Further
under the $H_{j+1}$-isomorphism $\hatf_{j+1}$, we know $T'$ is also
a right transversal of $H_{j+1}//J_{j+1}$.  Lastly under the 
$H_j$-isomorphism $f_{j,j+1}$, we know $\{(b,b') | b\in T\}$ is a collection
of trellis path segments of length 2 that form a right transversal of
$H_{[j,j+1]}//J_{[j,j+1]}$.

Now applying Figure \ref{fig4} repeatedly as we did with Figure \ref{fig3},
we can obtain the following extension of Theorem \ref{thmd}.

\begin{thm}
\label{thmc}
Fix $l$ such that $0<l\le k-m+1$.  There are two results:

(i) We can select a right transversal of $H_{[0,l]}//J_{[0,l]}$ such that
for $j=0,1,\ldots,l$, component $j$ forms a right transversal of $H_j//J_j$.
In particular we can choose the right transversal of $H_{[0,l]}//J_{[0,l]}$ 
so that component $j$ uses elements of $X_j\cap Y_{l-j}$
taken from right cosets of 
$$
(X_j\cap Y_{l-j})//D_j.
$$

(ii) Assume that $J_0\lhd H_0$.  Then $J_i\lhd H_i$, $i>0$, and
$J_{[0,l]}\lhd H_{[0,l]}$.
We can select a transversal of $H_{[0,l]}/J_{[0,l]}$ such that
for $j=0,1,\ldots,l$, component $j$ forms a transversal of $H_j/J_j$.
In particular we can choose the transversal of $H_{[0,l]}/J_{[0,l]}$ 
so that component $j$ uses elements of $X_j\cap Y_{l-j}$
taken from cosets of 
$$
(X_j\cap Y_{l-j})/D_j.
$$
\end{thm}

\vspace{3mm}
{\bf 5.  ENCODER FOR $\xj$ AND $\yk$}
\vspace{3mm}


Fix $j$ and $i$ such that $j\ge 0$, $i\ge 0$, and $j+i\le\ell$.
From Theorem \ref{thm17}, we know that we can find coset
representatives of  
\be
\label{qg1}
\frac{X_{j-1}(X_j\cap Y_i)}{X_{j-1}(X_j\cap Y_{i-1})}
\ee
using elements of $X_j\cap Y_i$.  Thus let $q_{j,i}$ be an 
element of $X_j\cap Y_i$ and $[q_{j,i}]$ be a transversal of (\ref{qg1}).
We can arrange the $[q_{j,i}]$ as shown below:
\be
\label{rep1}
\begin{array}{llllll}
  [q_{0,\ell}]   & [q_{1,\ell-1}] & [q_{2,\ell-2}] & \cdots & [q_{\ell-1,1}] & [q_{\ell,0}] \\
 {[q_{0,\ell-1}]}& [q_{1,\ell-2}] & [q_{2,\ell-3}] & \cdots & [q_{\ell-1,0}] & \\
  \vdots & \vdots & \vdots & \vdots && \\
 {[q_{0,2}]}  & [q_{1,1}] & [q_{2,0}] &&& \\
 {[q_{0,1}]}  & [q_{1,0}] &&&& \\
 {[q_{0,0}]}  &&&&&
\end{array}
\ee
Note that $[q_{0,0}]$ is just the elements of $X_0\cap Y_0$, i.e.,
$[q_{0,0}]=X_0\cap Y_0$.
It can be seen the array (\ref{rep1}) corresponds to the \csm\ (\ref{csm})
and $[q_{j,i}]$ corresponds to terms in the \csm.  The \csm\ gives a
{\it chain coset decomposition} of $B$, and the elements of (\ref{rep1})
form a {\it chain of coset representatives}.  Thus any element $b\in B$ can
be written using elements of (\ref{rep1}) as
\begin{multline}
\label{rep1a}
b=q_{0,0}q_{0,1}q_{0,2}\cdots q_{0,\ell}q_{1,0}\cdots q_{1,\ell-1}q_{2,0}\cdots \\
      \cdots q_{j,0}\cdots q_{j,i}\cdots q_{j,\ell-j}\cdots q_{\ell,0},
\end{multline}
where the terms in product $(q_{j,0}\cdots q_{j,i}\cdots q_{j,\ell-j})$ 
are coset representatives from the $j$-th column of (\ref{rep1}), $0\le j\le\ell$.

We now show how a similar result can be obtained using Theorem \ref{thmc}.
Fix integer $k$, $0<k\le\ell$.  Choose $m=1$.  Choose $l=k-m+1$; then $l=k$.
Then for $j$, $0\le j\le k$, we have
\begin{align*}
H_j &=X_{j-1}(X_j\cap Y_{k-j}), \\
J_j &=X_{j-1}(X_j\cap Y_{k-j-1}).
\end{align*}
A transversal of $H_{[0,k]}/J_{[0,k]}$ consists
of a trellis path segment of length $k+1$:
\be
\label{tps}
r_{0,k},r_{1,k-1},\ldots,r_{j,k-j},\ldots,r_{k,0}.
\ee
By Theorem \ref{thmc} we can select $r_{j,k-j}$ that is an element of 
$X_j\cap Y_{k-j}$ and a representative of
$H_j/J_j$,
\be
\label{qg2}
\frac{X_{j-1}(X_j\cap Y_{k-j})}{X_{j-1}(X_j\cap Y_{k-j-1})}.
\ee
Thus the transversals of $H_{[0,k]}/J_{[0,k]}$ give a set of transversals
$$
[r_{0,k}],[r_{1,k-1}],\ldots,[r_{j,k-j}],\ldots,[r_{k,0}],
$$
where $[r_{j,k-j}]$ is a transversal of $H_j/J_j$, for $j=0,1,\ldots,k$.
Since this is true for each $k$, $0<k\le\ell$, we can obtain the 
following matrix, similar to (\ref{rep1}), 
which gives a chain of coset representatives of $B$:
\be
\label{rep2}
\begin{array}{llllll}
  [r_{0,\ell}]   & [r_{1,\ell-1}] & [r_{2,\ell-2}] & \cdots & [r_{\ell-1,1}] & [r_{\ell,0}] \\
 {[r_{0,\ell-1}]}& [r_{1,\ell-2}] & [r_{2,\ell-3}] & \cdots & [r_{\ell-1,0}] & \\
  \vdots & \vdots & \vdots & \vdots && \\
 {[r_{0,2}]}  & [r_{1,1}] & [r_{2,0}] &&& \\
 {[r_{0,1}]}  & [r_{1,0}] &&&& \\
 {[r_{0,0}]}  &&&&&
\end{array}
\ee
We define $[r_{0,0}]$ to be $X_0\cap Y_0$.
Comparing (\ref{qg1}) and (\ref{qg2}), we see that representatives in $[r_{j,k-j}]$ 
in (\ref{rep2}) are in the same quotient group as representatives in $[q_{j,i}]$ 
in (\ref{rep1}) when $k-j=i$.  Therefore as in (\ref{rep1a}),
we can write $b\in B$, and thus any $b_t\in B_t$, as
\begin{multline}
\label{beqn}
b_t=r_{0,0}r_{0,1}r_{0,2}\cdots r_{0,\ell}r_{1,0}\cdots r_{1,\ell-1}r_{2,0}\cdots \\
      \cdots r_{j,0}\cdots r_{j,k-j}\cdots r_{j,\ell-j}\cdots r_{\ell,0},
\end{multline}
where the terms in product $(r_{j,0}\cdots r_{j,k-j}\cdots r_{j,\ell-j})$ 
are coset representatives from the $j$-th column of (\ref{rep2}), $0\le j\le\ell$.

Consider again a representative (\ref{tps}) in the transversal
of $H_{[0,k]}/J_{[0,k]}$.  Since $[r_{k,0}]\subset X_k\cap Y_0$,
any representative $r_{k,0}\in [r_{k,0}]$ merges to the identity state.  
Thus the components (\ref{tps}) 
of a transversal of $H_{[0,k]}/J_{[0,k]}$ can be extended with the 
identity element to form a path
$$
\ldots,\bone,\bone,r_{0,k},r_{1,k-1},\ldots,r_{j,k-j},\ldots,r_{k,0},\bone,\bone,\ldots
$$
in the trellis.  We call this path a {\it generator} $\bldg_{[0,k]}$; generator
$\bldg_{[0,k]}$ is the same as the identity path outside the time interval $[0,k]$.
Let $\Lambda_{[0,k]}$ be the set of generators formed by transversals of 
$H_{[0,k]}/J_{[0,k]}$, $0<k\le\ell$.  
For each $\bldg_{[0,k]}\in\Lambda_{[0,k]}$ and each time $t$, 
let $\bldg_{[t,t+k]}$ be the time
shift of $\bldg_{[0,k]}$; $\bldg_{[t,t+k]}$ is the same as the identity path 
outside the time interval $[t,t+k]$.  We say $\bldg_{[t,t+k]}$ is a 
{\it generator at time $t$}.  Let $\Lambda_{[t,t+k]}$ be the set of
generators $\bldg_{[t,t+k]}$, $0<k\le\ell$.
Let $\Lambda_{[0,0]}$ be the set of generators formed by elements of 
$[r_{0,0}]=X_0\cap Y_0$.  These are generators which are the same as
the identity path, except that time $0$ branch $b_0\in X_0\cap Y_0$.
Let $\Lambda_{[t,t]}$ be the time shift of these generators.

Fix time $t$.  We now show how to realize any element in (\ref{beqn}) at 
time $t$.  Clearly any $r_{0,0}$ can be realized at time $t$ by 
$\chi_t(\bldg_{[t,t]})$ for some $\bldg_{[t,t]}\in\Lambda_{[t,t]}$.
For any element $r_{j,k-j}$ in (\ref{beqn}), $0<k\le\ell$ and $0\le j\le k$,  
there is a $\bldg_{[t-j,t-j+k]}\in\Lambda_{[t-j,t-j+k]}$ such that 
$$
\chi_t(\bldg_{[t-j,t-j+k]})=r_{j,k-j}.
$$
We can rewrite this as
$$
\chi_t(\bldg_{[t-j,t+(k-j)]})=r_{j,k-j}.
$$
Using index $i$ in place of $k-j$, we can rewrite $b_t$ in (\ref{beqn}) as
\be
\label{enc}
b_t=\prod_{j=0}^\ell \left(\prod_{i=0}^{\ell-j} \chi_t(\bldg_{[t-j,t+i]})\right).
\ee
Thus we have shown any element $b_t\in B_t$ can be written using generators
selected at times in the interval $[t-\ell,t]$, i.e., generators
$$
\bldg_{[t-j,t+i]},\text{ for }j=0,\ldots,\ell,\text{ for }i=0,\ldots,\ell-j.
$$

An {\it encoder} of the group trellis is a finite state machine that,
given a sequence of inputs, can produce any path (any sequence of states
and branches) in the group trellis.
We assume the encoder has the same number of states as $B$.
An encoder can help to explain the structure
of a group trellis.  We give an encoder here which has a register structure
and uses shortest length sequences as in \cite{FT,LM}, but the encoder
is different.

We will define an encoder based on (\ref{enc}).
We can rewrite $b_t$ to separate out the $j=0$ term as
\begin{align}
\label{enca}
b_t= &\prod_{i=0}^\ell \chi_t(\bldg_{[t,t+i]}) \\
\label{encb}
     &\prod_{j=1}^\ell \left(\prod_{i=0}^{\ell-j} \chi_t(\bldg_{[t-j,t+i]})\right).
\end{align}
The term in (\ref{enca}) is just an element $x_t\in X_0$.
Note that $x_t$ is a function of generators at time $t$.  We will
think of $x_t$ as an {\it input} of the encoder.
The term in (\ref{encb}) is just a branch $\hatb_t\in B_t$; it
corresponds to the branch obtained when $x_t=\bone$.  Note that 
$\hatb_t$ is a function of generators at times $t'$, $t'<t$.
We can rewrite $b_t$ as $b_t=x_t\hatb_t$, where $\hatb_t\in B_t$
and 
$$
\hatb_t=\prod_{j=1}^\ell \left(\prod_{i=0}^{\ell-j} \chi_t(\bldg_{[t-j,t+i]})\right).
$$
Note that $\hatb_t$ can take on $|B/X_0|$ different values, and each value
is in a distinct coset of $B/X_0$.  Thus there is a one to one 
correspondence between the set of $\hatb_t$, $\{\hatb_t\}$, and $B/X_0$,
$\{\hatb_t\}\lra B/X_0$.  Thus we will think of $\hatb_t$ as a {\it state}
of the encoder.

We propose an encoder based on (\ref{enc}) which has the same form as
(\ref{enc}) for each time epoch.  Accordingly, the form of the encoder
for time $t+1$ is 
$$
b_{t+1}=\prod_{j=0}^\ell \prod_{i=0}^{\ell-j} \chi_{t+1}(\bldg_{[t+1-j,t+1+i]}').
$$
We can rewrite this to separate out the $j=0$ term:
\begin{align}
\label{enc1}
b_{t+1} = &\prod_{i=0}^\ell \chi_{t+1}(\bldg_{[t+1,t+1+i]}') \\
\label{enc2}
          &\prod_{j=1}^\ell \prod_{i=0}^{\ell-j} \chi_{t+1}(\bldg_{[t+1-j,t+1+i]}').
\end{align}
The term in (\ref{enc1}) uses $\ell+1$ new generators at time $t+1$,
$\bldg_{[t+1,t+1+i]}'$, for $i=0,\ldots,\ell$.  It is the new input
$x_{t+1}\in X_0$.  Let $\hatb_{t+1}$ be the term in (\ref{enc2}).
We see that $\hatb_{t+1}$ uses generators $\bldg_{[t+1-j,t+1+i]}'$
at times in the interval $[t,t-\ell+1]$.

We complete the specification of the encoder by requiring that the
generators in (\ref{enc2}), or $\hatb_{t+1}$, 
used by the encoder at time $t+1$ be the same
as the generators used by the encoder at time $t$.  In other words, we
require
$$
\bldg_{[t+1-j,t+1+i]}'=\bldg_{[t+1-j,t+1+i]}
$$
for $j=1,\ldots,\ell$, for $i=0,\ldots,\ell-j$.  Then
\be
\label{enc4}
\hatb_{t+1}=\prod_{j=1}^\ell \prod_{i=0}^{\ell-j} \chi_{t+1}(\bldg_{[t+1-j,t+1+i]}).
\ee
The encoder output at time $t+1$ is $b_{t+1}=x_{t+1}\hatb_{t+1}$.
Thus the encoder uses a sliding block encoding of the past, given by $\hatb_{t+1}$,
with new inputs at each time epoch, given by $x_{t+1}$.
With this definition of the encoder, note that for any input $x_{t+1}$
to the encoder, $(b_t,b_{t+1})$ is a trellis path segment of length two.
In particular, for $x_{t+1}=\bone$, we have $(b_t,\hatb_{t+1})$ is a trellis 
path segment of length two.

We now verify that given an arbitrary path $\bldc^*$ in the group trellis,
the encoder can track it perfectly, i.e., produce the same path.
First we give a useful lemma.  We think of the encoder as an estimator.

\begin{lem}
\label{lem10}
Let $\bldc$ and $\bldc^*$ be two paths in the group trellis $C$:
\begin{align*}
\bldc   &=\ldots,b_t,b_{t+1},b_{t+2},\ldots \\
\bldc^* &=\ldots,b_t^*,b_{t+1}^*,b_{t+2}^*,\ldots .
\end{align*}
The two trellis paths are arbitrary except that at time $t$,
$b_t=b_t^*$.  Then at time $t+1$, $b_{t+1}$ is in the
same coset of $B/X_0$ as $b_{t+1}^*$, i.e., $b_{t+1}^*\in X_0 b_{t+1}$.  
At time $t+2$, $b_{t+2}$ is in the same coset of $B/X_1$ as $b_{t+2}^*$.  
In general at time $t+j$, $j\le\ell$,
$b_{t+j}$ is in the same coset of $B/X_{j-1}$ as $b_{t+j}^*$.
For time $t+j$, $j>\ell$, $b_{t+j}$ is in the same coset of $B/X_\ell$ as $b_{t+j}^*$;
this coset is just $B$.  In a similar way, at time $t-1$, $b_{t-1}$ is in
the same coset of $B/Y_0$ as $b_{t-1}^*$.
\end{lem}

\begin{prf}
We have $\caln(b_t)=\caln(b_t^*)$, so $b_{t+1}$ and $b_{t+1}^*$ are in
the same coset of $X_0$.  The rest of the proof is analogous.
\end{prf}

We can think of $\bldc$ as a path which estimates $\bldc^*$.
The lemma shows how the estimation degrades as time goes on when it is
perfect at $t=0$, for two otherwise arbitrary paths.  Note that the
conclusion of the lemma is unchanged if, instead of one path
$\bldc$, we use a finite number of paths to estimate $\bldc^*$.

We now show that the encoder can track arbitrary path $\bldc^*$.
Fix time $t$.  The branch in $\bldc^*$ at time $t$ is $b_t^*$.
First we show that the encoder can produce the initial condition $b_t^*$.
But we already know the encoder can produce any branch $b_t=b_t^*$
(see (\ref{enc})).  
Thus at time $t$, the encoder can produce the branch $b_t^*$ in $\bldc^*$.

Now we show the encoder can track path $\bldc^*$.  
First we show the encoder can track $\bldc^*$ at time $t+1$.  
First find $\hatb_{t+1}$ as given in (\ref{enc4}).  
We know $(b_t,\hatb_{t+1})$ is a trellis 
path segment of length two.
Therefore we know from Lemma \ref{lem10} that $b_{t+1}^*\in X_0 \hatb_{t+1}$.  
Thus there exists $\hatx_{t+1}\in X_0$ such that $b_{t+1}^*=\hatx_{t+1}\hatb_{t+1}$.  
Now select $\ell+1$ new generators at time $t+1$,
$\bldhatg_{[t+1,t+1+i]}$, for $i=0,\ldots,\ell$, such that
$$
\prod_{i=0}^\ell \chi_{t+1}(\bldhatg_{[t+1,t+1+i]})=\hatx_{t+1}.
$$
We know this is possible since $\hatx_{t+1}\in X_0$ and the first column
of (\ref{rep2}) is a coset decomposition of $X_0$.
With input $\hatx_{t+1}$ at time $t+1$, the encoder output
is given by $\hatx_{t+1}\hatb_{t+1}=b_{t+1}^*$.
Thus we see that with proper input $\hatx_{t+1}$, the encoder can track path $\bldc^*$
at time $t+1$.  We can repeat this same argument for succeeding time epochs.

We can observe several features of the encoder.  The term in the parentheses of 
(\ref{enc}) is some function of $t-j$, say $h_{t-j}$.  
Then $b_t=\prod_{j=0}^\ell h_{t-j}$.  Thus the encoder has the form
of a convolution, reminiscent of a linear system.
The Forney-Trott encoder \cite{FT} and 
Loeliger-Mittelholzer encoder \cite{LM} do not have the convolution
analogy.

Note that if we apply input $x_t=\chi_t(\bldg_{[t,t+k]})$
at time $t$ to an encoder in the identity state, the output is just
$$
\chi_{t+n}(\bldg_{[t,t+k]}),\text{ for }n=0,1,\ldots,\infty.
$$
Thus the output response of the encoder is the input generator.
This is reminiscent of the impulse response of a linear system.

In view of Figure \ref{fig2}, the \csm\ (\ref{csm}), and (\ref{rep2}), 
a natural interpretation of the encoder is shown in Figure
\ref{fig5}.  The encoder consists of $\ell+1$ registers, where the $j^{\rm th}$
register, $j=0,1,\ldots,\ell$, contains the information (coset representatives) 
in quotient group $X_j/X_{j-1}$.  At the next time epoch, the register
containing $X_j/X_{j-1}$ is shifted or ``shoved'' into the register 
containing  $X_{j+1}/X_j$.  The encoder shown in Figure \ref{fig5}
is somewhat different from the classic interpretation of a shift
register.

\begin{figure}[h]
\centering
\vspace{3ex}

\begin{picture}(120,115)(0,-10)

\put(0,0){\framebox(20,100)}
\put(40,20){\framebox(20,80)}
\put(60,40){\framebox(20,60)}
\put(80,60){\framebox(20,40)}
\put(100,80){\framebox(20,20)}

\put(7.5,30){\shortstack{i\\n\\p\\u\\t}}

\put(20,50){\vector(1,0){20}}
\put(40,110){\vector(1,0){80}}
\put(120,110){\vector(-1,0){80}}

\put(80,115){\makebox(0,0)[b]{$\ell$}}

\put(10,-10){\makebox(0,0){$\frac{X_0}{\bone}$}}
\put(50,-10){\makebox(0,0){$\frac{X_1}{X_0}$}}
\put(70,-10){\makebox(0,0){$\frac{X_2}{X_1}$}}
\put(90,-10){\makebox(0,0){$\cdots$}}
\put(110,-10){\makebox(0,0){$\frac{X_\ell}{X_{\ell-1}}$}}

\end{picture}

\caption{Shift/shove register.}
\label{fig5}

\end{figure}
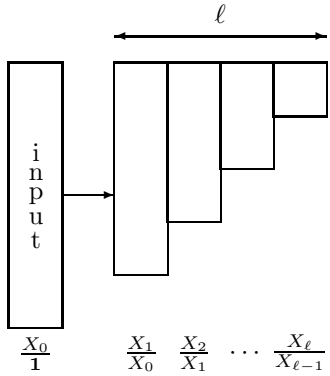

Any arbitrary group $G$ has a {\it chain coset decomposition}.
Any element $g\in G$ can be specified by the set $R$ of 
{\it chain coset representatives}.

\begin{thm}
Let $G$ be an arbitrary group.  Let $G$ have two chain coset decompositions
and two sets of chain coset representatives, $R$ and $R'$, respectively.
Then $|R|=|R'|$.
\end{thm}

\begin{prf}
Any chain coset decomposition of $G$ can be refined into a composition
chain.  Then the two chain coset decompositions with sets of chain coset
representatives $R$ and $R'$ can be refined into two composition chains
with sets of chain coset representatives $R_c$ and $R_c'$, respectively.
But we must have $|R|=|R_c|$ and $|R'|=|R_c'|$.  But by the 
Jordan-H\"{o}lder theorem, any two composition chains are equivalent.
Thus $|R_c|=|R_c'|$ and so $|R|=|R'|$.
\end{prf}

Thus we can associate with any group $G$ a unique number $\eta_G$,
such that $\eta_G$ elements of $G$ are necessary and sufficient to
specify an arbitrary element of $G$, irrespective of the chain coset
decomposition used.

At each time $t$, our encoder outputs an element $b_t\in B$.  We say the
encoder is {\it minimal} because at each time epoch $t$, it
uses $\eta_B$ elements to specify $b_t$.  This definition of minimality
is different from the approach in \cite{FT,LM,LFMT}.  However the
number of elements and number of states used in the encoder here and
the encoders in \cite{FT,LM} are the same.

\vspace{3mm}
{\bf 6.  COMPOSITION SERIES OF $B$}
\vspace{3mm}

\newcommand{\bigx}[3]{{\mathbf{X}}_{#1,#2}^{(#3)}}
\newcommand{\bigy}[3]{{\mathbf{Y}}_{#1,#2}^{(#3)}}

For each branch $b\in B$, we define the {\it previous branch set} $\calp(b)$
to be the set of branches that can precede $b$ at the previous
time epoch in valid trellis paths.  
In other words, branch $e\in\calp(b)$ \ifof\ $b^-=e^+$.

For a set $U\subset B$, define the set $\calp(U)$ to be
the union $\cup_{b\in U} \calp(b)$.  The set $\calp(U)$ always
consists of cosets of $Y_0$.  Note that $\calp(Y_j)=Y_{j+1}$.

\begin{prop}
If $b^-=e^+$, the previous
branch set $\caln(b)$ of a branch $b$ in $B$ is the coset $Y_0e$ in $B$.
If $b^-=e^+$, the previous branch set $\calp(b)$ of a branch $b$ in 
$Y_j$ is the coset $Y_0e$ in $Y_{j+1}$.  
\end{prop}

Note that $\calp(X_{j+1})\cap X_j=X_j$.
For a set $U'\subset X_{j+1}$, define $\calp_j(U')=\calp(U')\cap X_j$.

\begin{prop}
\label{prop411}
If $U'\subset X_{j+1}$ is a group, then $\calp_j(U')$
is a group.  If $U'\subset X_{j+1}$ is a set, then $\calp_j(U')$
consists of cosets of $X_j\cap Y_0$.
If set $U\subset X_j$ consists of cosets of $X_j\cap Y_0$, then
$U=\calp_j(\caln(U))$.  
\end{prop}

\begin{prf}
If $U'$ is a group, then $\calp(U')$ is a group so $\calp(U')\cap X_j$
is a group.  This proves the first statement.

Assume $U'\subset X_{j+1}$ is a set.  Let state $s\in (\calp_j(U'))^+$.
Then $s\in (\calp(U')\cap X_j)^+=(\calp(U'))^+\cap X_j^+$.
Therefore $s\in (\calp(U'))^+$ and $s\in X_j^+$.  But branches
in $\calp(U')$ that merge to $s$ are a coset of $Y_0$, and branches
in $X_j$ that merge to $s$ are a coset of $X_j\cap Y_0$ (by
Proposition \ref{prop1a}).  Then branches in $\calp(U')\cap X_j$
that merge to $s$ are a coset of $X_j\cap Y_0$.  This proves the
second statement.

The third statement follows from the second.
\end{prf}

The following proposition is similar to (iv) of Proposition \ref{prop40}.

\begin{prop}
\label{prop41}
Assume $J',H'$ are subgroups of $X_{j+1}$ and $X_0\le J'\le H'$.
Then there exist subsets $J,H\subset X_j$ such that $\caln(J)=J'$ and 
$\caln(H)=H'$.  In particular, we can choose $J=\calp_j(J')$ and 
$H=\calp_j(H')$.  Then $J,H$ are groups and $J\le H$.  
And if $J'\lhd H'$, then $J\lhd H$.
\end{prop}

We want to find a refinement of the \sm\ (\ref{csm}), which we denote by
$\{\bigx{j}{k}{\sigma}\}$.  The refinement has terms $\bigx{j}{k}{\sigma}$, 
for $j\ge 0$, $k\ge -1$ such that $j+k<\ell$, for $\sigma$ such that
$0\le \sigma\le\sigma_{j+k}$, as shown here:

\begin{align}
\label{seriesaa}
&X_{-1}(\bone) \\
\label{seriesaaa}
=&\bigx{0}{-1}{0}\subset\bigx{0}{-1}{1}\subset\ldots\subset\bigx{0}{-1}{\sigma_{-1}}  \\
\nonumber
=&\bigx{0}{0}{0}\subset\bigx{0}{0}{1}\subset\ldots\subset\bigx{0}{0}{\sigma_0}  \\
\nonumber
=&\bigx{0}{1}{0}\subset\bigx{0}{1}{1}\subset\ldots\subset\bigx{0}{1}{\sigma_1}  \\
\nonumber
&\cdots \\
\label{seriesb}
=&\bigx{0}{\ell-1}{0}\subset\bigx{0}{\ell-1}{1}\subset\ldots\subset\bigx{0}{\ell-1}{\sigma_{\ell-1}}  \\
\nonumber
=&X_0(\bone) \\
\label{seriesd}
=&\bigx{1}{-1}{0}\subset\bigx{1}{-1}{1}\subset\ldots\subset\bigx{1}{-1}{\sigma_{0}}  \\
\nonumber
=&\bigx{1}{0}{0}\subset\bigx{1}{0}{1}\subset\ldots\subset\bigx{1}{0}{\sigma_{1}}  \\
\nonumber
&\cdots \\
\label{seriese}
=&\bigx{1}{\ell-2}{0}\subset\bigx{1}{\ell-2}{1}\subset\ldots\subset\bigx{1}{\ell-2}{\sigma_{\ell-1}}  \\
\nonumber
=&X_1(\bone) \\
\nonumber
&\cdots \\
\nonumber
=&\bigx{j}{-1}{0}\subset\bigx{j}{-1}{1}\subset\ldots\subset\bigx{j}{-1}{\sigma_{j-1}}  \\
\nonumber
&\cdots \\
\label{eqn34}
=&\bigx{j}{k}{0}\subset\bigx{j}{k}{1}\subset\ldots\subset\bigx{j}{k}{\sigma_{j+k}}  \\
\nonumber
=&\bigx{j}{k+1}{0}\subset\bigx{j}{k+1}{1}\subset\ldots\subset\bigx{j}{k+1}{\sigma_{j+k+1}}  \\
\nonumber
&\cdots \\
\nonumber
=&\bigx{j}{\ell-j-1}{0}\subset\bigx{j}{\ell-j-1}{1}\subset\ldots\subset\bigx{j}{\ell-j-1}{\sigma_{\ell-1}}  \\
\nonumber
=&X_j(\bone) \\
\nonumber
=&\bigx{j+1}{-1}{0}\subset\bigx{j+1}{-1}{1}\subset\ldots\subset\bigx{j+1}{-1}{\sigma_{j}}  \\
\nonumber
&\cdots \\
\nonumber
=&\bigx{\ell}{-1}{0}\subset\bigx{\ell}{-1}{1}\subset\ldots\subset\bigx{\ell}{-1}{\sigma_{\ell-1}}  \\
\label{seriesc}
=&X_\ell(\bone).
\end{align}

For $j\ge 0$, $k\ge -1$, $j+k<\ell$, we let 
\be
\label{refer1}   
\bigx{j}{k}{\sigma_{j+k}}{\stackrel{\rm def}{=}}X_{j-1}(X_j\cap Y_{k+1}).
\ee
Then (\ref{seriesaa})-(\ref{seriesc}) contains all terms of the \sm\
(\ref{csm}).  And after a description of the remaining terms,
(\ref{seriesaa})-(\ref{seriesc}) will be a refinement of the \sm\
(\ref{csm}).  
Since we assume the \sm\ is \ellctl, it follows that 
for $j\ge 0$, $k\ge -1$, $j+k=\ell-1$, 
\be
\label{refer2}
\bigx{j}{\ell-j-1}{\sigma_{\ell-1}}=X_{j-1}(X_j\cap Y_{\ell-j})=X_j,
\ee
as shown in (\ref{seriesaa})-(\ref{seriesc}).  

Note that the construction uses $\bigx{j}{k}{\sigma_{j+k}}=\bigx{j}{k+1}{0}$ 
for $j\ge 0$, $k\ge -1$, $j+k<\ell-1$, 
and $\bigx{j}{\ell-j-1}{\sigma_{\ell-1}}=\bigx{j+1}{-1}{0}$
for $j\ge 0$, $k\ge -1$, $j+k=\ell-1$, $j\ne\ell$. 
It follows that 
$$   
\bigx{j}{k}{0}=X_{j-1}(X_j\cap Y_k)
$$
for $j\ge 0$, $k\ge -1$, $j+k<\ell$, and 
\be
\label{refer3}
\bigx{j}{-1}{0}=X_{j-1}
\ee
for $0\le j\le\ell$.

In general the subchain
\be
\label{schn34}
\bigx{j}{k}{0}\subset\bigx{j}{k}{1}\subset\ldots\subset\bigx{j}{k}{\sigma}
\subset\ldots\subset\bigx{j}{k}{\sigma_{j+k}}
\ee
in (\ref{eqn34}) is a refinement of the subchain 
$X_{j-1}(X_j\cap Y_k)\subset X_{j-1}(X_j\cap Y_{k+1})$ in the \sm\ (\ref{csm}).
The groups in subchain (\ref{schn34}) are indexed by $\sigma$.
We show in Theorem \ref{thm41} that we can find a refinement which is
a composition chain in which the number of groups in subchain 
(\ref{schn34}) just depends on the sum $j+k$; then index $\sigma$
runs from $0$ to parameter $\sigma_{j+k}$ as shown in (\ref{eqn34}).
In fact Theorem \ref{thm41} shows that for $k\ge 0$,
$$
\caln(\bigx{j}{k}{\sigma})=\bigx{j+1}{k-1}{\sigma},
$$
for $\sigma=0,1,\ldots,\sigma_{j+k}$.

\begin{thm}
\label{thm41}
Let $B$ be an \ellctl\ group trellis section with normal chains $\xj$ and
$\yk$.  There is a refinement $\{\bigx{j}{k}{\sigma}\}$ of the \sm\ (\ref{csm}), 
given by (\ref{seriesaa})-(\ref{seriesc}), which is a 
composition chain of $\xj$ (and $B$).  The terms $\bigx{j}{k}{\sigma}$ 
exist for $j\ge 0$, $k\ge -1$ such that $j+k<\ell$, for 
$0\le \sigma\le\sigma_{j+k}$.  For $j\ge 0$, $k\ge 0$ such that $j+k<\ell$, for 
$0\le \sigma\le\sigma_{j+k}$, if
$\bigx{j}{k}{\sigma}$ is a group in the composition chain, 
then $\caln(\bigx{j}{k}{\sigma})$ is the group
$\bigx{j+1}{k-1}{\sigma}$ in the composition chain.  Lastly,
fix $k$ such that $0\le k<\ell$.  Then for $j$ such that 
$0\le j\le k+1$, and $\sigma$ such that $0<\sigma\le\sigma_k$, 
we have isomorphisms
\be
\label{eot41}
\frac{\bigx{0}{k}{\sigma}}{\bigx{0}{k}{\sigma-1}}
\simeq\frac{\caln^j(\bigx{0}{k}{\sigma})}{\caln^j(\bigx{0}{k}{\sigma-1})}  
=\frac{\bigx{j}{k-j}{\sigma}}{\bigx{j}{k-j}{\sigma-1}}.
\ee
\end{thm}

\begin{prf}
First we find a refinement of 
$[\bone,X_{-1}(X_0\cap Y_\ell)]=[\bigx{0}{-1}{0},\bigx{0}{\ell-1}{\sigma_{\ell-1}}]$, 
the first column of (\ref{csm}) 
(for a normal chain containing groups $G,H$, the notation $[G,H]$
means the subchain starting with group $G$ and ending with group $H$, i.e.,
the subchain with groups in the interval $[G,H]$).
We select the refinement so it is a composition chain; this refinement 
is given in (\ref{seriesaaa})-(\ref{seriesb}).  In general the groups
in the composition chain of subchain $[\bigx{0}{k}{0},\bigx{0}{k}{\sigma_k}]$
are indexed by $\sigma$, $\sigma=0,1,\ldots,\sigma_k$, for 
$k=-1,0,\ldots,\ell-1$,
\be
\label{schn40}
\bigx{0}{k}{0}\lhd\bigx{0}{k}{1}\lhd\ldots
\lhd\bigx{0}{k}{\sigma-1}\lhd\bigx{0}{k}{\sigma}
\lhd\ldots\lhd\bigx{0}{k}{\sigma_k}.
\ee

Next we show how to find a refinement of 
$[X_0,X_0(X_1\cap Y_{\ell-1})]=[\bigx{1}{-1}{0},\bigx{1}{\ell-2}{\sigma_{\ell-1}}]$, 
the second column of (\ref{csm}).
The refinement of $[\bigx{1}{-1}{0},\bigx{1}{\ell-2}{\sigma_{\ell-1}}]$ is found by taking
the refinement of $[\bigx{0}{0}{0},\bigx{0}{\ell-1}{\sigma_{\ell-1}}]$ previously found
and applying $\caln$ to each of the terms.  For example for the subchain
$[\bigx{0}{k}{0},\bigx{0}{k}{\sigma_k}]$, $0\le k\le\ell-1$, of 
$[\bigx{0}{0}{0},\bigx{0}{\ell-1}{\sigma_{\ell-1}}]$ shown in (\ref{schn40}), the
corresponding subchain of $[\bigx{1}{-1}{0},\bigx{1}{\ell-2}{\sigma_{\ell-1}}]$ is
\begin{multline}
\label{schn41}
\caln(\bigx{0}{k}{0})\subset\caln(\bigx{0}{k}{1})\subset\ldots
\subset\caln(\bigx{0}{k}{\sigma-1})\subset\caln(\bigx{0}{k}{\sigma})\subset \\
\ldots\subset\caln(\bigx{0}{k}{\sigma_k}).
\end{multline}
Using Proposition \ref{prop10} and Proposition \ref{prop14}, we have 
$\caln(\bigx{0}{k}{0})=\bigx{1}{k-1}{0}$ and 
$\caln(\bigx{0}{k}{\sigma_k})=\bigx{1}{k-1}{\sigma_k}$.
Then (\ref{schn41}) becomes
\begin{multline}
\label{schn42}
\bigx{1}{k-1}{0}\subset\caln(\bigx{0}{k}{1})\subset\ldots
\subset\caln(\bigx{0}{k}{\sigma-1})\subset\caln(\bigx{0}{k}{\sigma})\subset \\
\ldots\subset\bigx{1}{k-1}{\sigma_k}.
\end{multline}
Since $\bigx{0}{k}{\sigma-1}\lhd\bigx{0}{k}{\sigma}$, we have by 
Proposition \ref{prop40} that
$\caln(\bigx{0}{k}{\sigma-1})\lhd\caln(\bigx{0}{k}{\sigma})$.
Then (\ref{schn42}) becomes
\be
\label{schn43}
\bigx{1}{k-1}{0}\lhd\caln(\bigx{0}{k}{1})\lhd\ldots
\lhd\caln(\bigx{0}{k}{\sigma-1})\lhd\caln(\bigx{0}{k}{\sigma})
\lhd\ldots\lhd\bigx{1}{k-1}{\sigma_k},
\ee
and so we have found a refinement of $[\bigx{1}{-1}{0},\bigx{1}{\ell-2}{\sigma_{\ell-1}}]$.

Next we show the refinement of $[\bigx{1}{-1}{0},\bigx{1}{\ell-2}{\sigma_{\ell-1}}]$
is a composition chain.  If the refinement is not a composition chain,
then we can insert a group, say $J'$, in the chain which gives a nontrivial
refinement.  Without loss of generality assume we can insert $J'$ in
(\ref{schn43}) so that 
\be
\label{eq44}
\caln(\bigx{0}{k}{\sigma-1})\lhd J'\lhd\caln(\bigx{0}{k}{\sigma})
\ee
is a nontrivial chain.  Since 
$\caln(\bigx{0}{k}{\sigma})\subset\bigx{1}{\ell-2}{\sigma_{\ell-1}}=X_1$, and
$X_0=\bigx{1}{-1}{0}\subset\caln(\bigx{0}{k}{\sigma-1})$, we can apply
Proposition \ref{prop41}.  Then there exists groups
$\calp_0(\caln(\bigx{0}{k}{\sigma-1}))$, $J=\calp_0(J')$, and
$\calp_0(\caln(\bigx{0}{k}{\sigma}))$ in $X_0$ such that
\be
\label{eq44a}
\calp_0(\caln(\bigx{0}{k}{\sigma-1}))\lhd J\lhd\calp_0(\caln(\bigx{0}{k}{\sigma})).
\ee
Since $X_0\cap Y_0=\bigx{0}{0}{0}\subset\bigx{0}{k}{\sigma-1}$, 
by Proposition \ref{prop411}, (\ref{eq44a}) is the same as
\be
\label{eq45}
\bigx{0}{k}{\sigma-1}\lhd J\lhd\bigx{0}{k}{\sigma}.
\ee
It is clear (\ref{eq45}) is a nontrivial chain since (\ref{eq44})
is a nontrivial chain.  Thus we have found a nontrivial refinement
of the composition chain of $[\bigx{0}{k}{0},\bigx{0}{k}{\sigma_k}]$,
a contradiction.  This shows the above procedure gives a 
composition chain of $[\bigx{1}{-1}{0},\bigx{1}{\ell-2}{\sigma_{\ell-1}}]$.
Finally, define terms of the composition chain $\{\bigx{j}{k}{\sigma}\}$ 
in interval $[\bigx{1}{-1}{0},\bigx{1}{\ell-2}{\sigma_{\ell-1}}]$ by
$$
\bigx{1}{k-1}{\sigma}{\stackrel{\rm def}{=}}\caln(\bigx{0}{k}{\sigma}),
$$
for $0<\sigma<\sigma_k$, $-1\le k\le\ell-2$.  This gives the 
refinement in (\ref{seriesd})-(\ref{seriese}).

The theorem assertion that $\caln(\bigx{0}{k}{\sigma})=
\bigx{1}{k-1}{\sigma}$ follows by construction.  Note that
$\bigx{0}{0}{0}=X_{-1}^*$.  Therefore from (\ref{cut2}) of Theorem \ref{thm22},
$\caln$ defines an isomorphism
$$
\frac{\bigx{0}{\ell-1}{\sigma_{\ell-1}}}{\bigx{0}{0}{0}}
\simeq\frac{\caln(\bigx{0}{\ell-1}{\sigma_{\ell-1}})}{\caln(\bigx{0}{0}{0})}.
$$
Therefore the construction procedure applying $\caln$ to the terms
of $[\bigx{0}{0}{0},\bigx{0}{\ell-1}{\sigma_{\ell-1}}]$ gives (\ref{eot41}) for $j=1$.

Next using the same approach we find a composition chain of 
$[\bigx{2}{-1}{0},\bigx{2}{\ell-3}{\sigma_{\ell-1}}]$.  Continuing in this way
gives a composition chain $\{\bigx{j}{k}{\sigma}\}$ of (\ref{csm}),
and in like manner the other theorem assertions hold.
\end{prf}

\begin{thm}
Let $B$ be an \ellctl\ group trellis section with normal chains $\xj$ and
$\yk$.  Then $B$ is solvable \ifof\ $X_0$ is solvable.
\end{thm}

\begin{prf}
If $G$ is solvable, then every subgroup is solvable, so $X_0$ is
solvable.  For the converse result, assume that $X_0$ is solvable.
Then the composition chain $[\bone,X_0]$ constructed in Theorem \ref{thm41} is solvable.
By Theorem \ref{thm41}, this means the chain in $[X_j,X_{j+1}]$ is
solvable for $0\le j<\ell$.  Then the entire chain $[\bone,X_\ell]$ is 
solvable and $B$ is solvable.
\end{prf}

\vspace{3mm}
{\bf 7.  ENCODER FOR $\xj$ AND $\{\bigy{k}{\rho}{\sigma}\}$}
\vspace{3mm}

\newcommand{\cupeq}{\cup\!\shortparallel}

It is clear that $\yk$ is the dual of $\xj$, and $\calp$ is the dual
of $\caln$.  Then we can obtain the following dual result to Theorem \ref{thm41},
giving the dual composition chain $\{\bigy{k}{j}{\sigma}\}$ of 
$\{\bigx{j}{k}{\sigma}\}$.

For a set $U\subset B$ and integer $i>0$, define $\calp^i(U)$
to be the $i$-fold composition $\calp^i(U)=\calp\circ\calp\circ\cdots\circ\calp(U)$.
For $i=0$, define $\calp^i(U)=\calp^0(U)$ to be just $U$.

\begin{thm}
\label{thm42}
Let $B$ be an \ellctl\ group trellis section with normal chains $\xj$ and
$\yk$.  There is a refinement $\{\bigy{k}{j}{\sigma}\}$ of the \sm\ 
of $\yk$ and $\xj$ which is a composition chain of $\yk$ (and $B$).  
The terms $\bigy{k}{j}{\sigma}$ 
exist for $k\ge 0$, $j\ge -1$ such that $j+k<\ell$, for 
$0\le \sigma\le\sigma_{j+k}$.  For $k\ge 0$, $j\ge 0$ such that $j+k<\ell$, for 
$0\le \sigma\le\sigma_{j+k}$, if
$\bigy{k}{j}{\sigma}$ is a group in the composition chain, 
then $\calp(\bigy{k}{j}{\sigma})$ is the group
$\bigy{k+1}{j-1}{\sigma}$ in the composition chain.  
Lastly, fix $j$ such that $0\le j<\ell$.  Then for $k$ such that 
$0\le k\le j+1$, and $\sigma$ such that $0<\sigma\le\sigma_j$, 
we have isomorphisms
\be
\label{eot42}
\frac{\bigy{0}{j}{\sigma}}{\bigy{0}{j}{\sigma-1}}
\simeq\frac{\calp^k(\bigy{0}{j}{\sigma})}{\calp^k(\bigy{0}{j}{\sigma-1})}  
=\frac{\bigy{k}{j-k}{\sigma}}{\bigy{k}{j-k}{\sigma-1}}.
\ee
\end{thm}

Find a composition chain $\{\bigy{k}{\rho}{\sigma}\}$ of $\yk$, 
where we have used index $\rho$ in place of index $j$.
Insert the composition chain $\{\bigy{k}{\rho}{\sigma}\}$ into the normal series
$\xj$, and vice versa.  This gives a refinement of $\xj$ and a refinement
of $\{\bigy{k}{\rho}{\sigma}\}$.  The two refinements are equivalent
by the Schreier refinement theorem.  Since the refinement of 
$\{\bigy{k}{\rho}{\sigma}\}$ is a composition series, 
we know both refinements are composition series.

The refinement of $\xj$ contains terms of the form 
$$
X_{j-1}(X_j\cap\bigy{k}{\rho}{\sigma})
$$
for $0\le j\le\ell$, for $k\ge 0$, $\rho\ge -1$ such that 
$k+\rho<\ell$, for $0\le\sigma\le\sigma_{k+\rho}$.  
We can think of the refinement or composition series of $\xj$ 
as a 4-dimensional array with indices $j$, $k$, $\rho$, and $\sigma$.  
We call this the {\it Schreier array form} of 
$\xj$ and $\{\bigy{k}{\rho}{\sigma}\}$.  

We think of a Schreier array as composed of pages.
For fixed $j$ such that $0\le j\le\ell$, 
and fixed $k$ such that $0\le k\le\ell$,
{\it page} $\Omega_{j,k}$ of the Schreier array is the set containing
all terms of the form
$$
X_{j-1}(X_j\cap\bigy{k}{\rho}{\sigma})
$$
for $\rho\ge -1$ such that $k+\rho<\ell$, for $\sigma$ 
such that $0\le\sigma\le\sigma_{k+\rho}$.  
We can write page $\Omega_{j,k}$ as a subchain of groups   

\begin{align}
\nonumber
& X_{j-1}(X_j\cap\bigy{k}{-1}{0})\subset\cdots\subset X_{j-1}(X_j\cap\bigy{k}{-1}{\sigma_{k-1}})  \\
\nonumber
\subset & X_{j-1}(X_j\cap\bigy{k}{0}{0})\subset\cdots\subset X_{j-1}(X_j\cap\bigy{k}{0}{\sigma_k})  \\
\label{sbchn}
& \cdots  \\
\nonumber
\subset & X_{j-1}(X_j\cap\bigy{k}{\ell-k-1}{0})\subset\cdots\subset X_{j-1}(X_j\cap\bigy{k}{\ell-k-1}{\sigma_{\ell-1}}).
\end{align}
Using the dual result to (\ref{refer3}), we have 
$\bigy{k}{-1}{0}=Y_{k-1}$, and so the first term of
subchain (\ref{sbchn}) is $X_{j-1}(X_j\cap Y_{k-1})$.
Using the dual result to (\ref{refer2}), we have 
$\bigy{k}{\ell-k-1}{\sigma_{\ell-1}}=Y_k$,
and so the last term of subchain (\ref{sbchn}) is
$X_{j-1}(X_j\cap Y_k)$.

Since the \sm\ of $\xj$ and $\yk$ in (\ref{sm}) is \ellctl, we know that
for $0\le j\le\ell$, $0\le k\le\ell$, and $j+k\ge\ell$,
$X_{j-1}(X_j\cap Y_k)=X_j$.  Then using (\ref{sbchn}) and its
endpoint conditions, for $0\le j\le\ell$, $0\le k\le\ell$, and $j+k>\ell$,
\be
\label{bndry}
\Omega_{j,k}=X_j.
\ee
Eliminating the trivially redundant pages
satisfying (\ref{bndry}), we can arrange the remaining 
pages into an inclusion chain of sets, as shown in (\ref{sac}):

\be
\label{sac}
\begin{array}{cccccc}
  \cupeq & \cupeq & \cupeq && \cupeq & \\

  \Omega_{0,\ell} & \Omega_{1,\ell-1} & \Omega_{2,\ell-2} & \cdots & \Omega_{\ell-1,1} & \Omega_{\ell,0} \\

  \cupeq & \cupeq & \cupeq && \cupeq & \\

  \Omega_{0,\ell-1} & \Omega_{1,\ell-2} & \Omega_{2,\ell-3} & \cdots & \Omega_{\ell-1,0} \\
 
  \cupeq & \cupeq & \cupeq &&& \\

  \cdots & \cdots & \cdots &&& \\

  \cupeq & \cupeq & \cupeq &&& \\

  \Omega_{0,2} & \Omega_{1,1} & \Omega_{2,0} &&& \\

  \cupeq & \cupeq &&&& \\
   
  \Omega_{0,1} & \Omega_{1,0} &&&& \\

  \cupeq &&&&& \\

  \Omega_{0,0} &&&&&
\end{array}
\ee
The indices $j,k$ of $\Omega_{j,k}$ in (\ref{sac}) satisfy
$j\ge 0$, $k\ge 0$, and $j+k\le\ell$.

The notation
$$
\begin{array}{c}
  V \\
  \cupeq \\
  U
\end{array}
$$
(also written $U\cupeq\! V$) means set $U$ is contained 
in $V$, and sets $U$ and $V$ share a common element.
For $j\ge 0$, $k\ge 0$ such that $j+k<\ell$, the common element
in $\Omega_{j,k}\cupeq\Omega_{j,k+1}$ is $X_{j-1}(X_j\cap Y_k)$.
For $j\ge 0$, $k\ge 0$, $j+k=\ell$, $j\ne\ell$, the common element
in $\Omega_{j,\ell-j}\cupeq\Omega_{j+1,0}$ is 
$X_{j-1}(X_j\cap Y_{\ell-j})=X_j$.
 
If we arrange the elements in each page into a 
subchain as in (\ref{sbchn}), then (\ref{sac}) represents a chain of groups.  
Regarded as a chain of groups, the page matrix (\ref{sac})
contains all terms in the \ellctl\ Schreier matrix (\ref{csm}), i.e., the terms
$X_{j-1}(X_j\cap Y_k)$ for $j\ge 0$, $k\ge -1$, $j+k\le\ell$.
Therefore as a chain of groups, the page matrix (\ref{sac}) is a refinement of 
the Schreier matrix (\ref{csm}).  We now show groups in 
(\ref{sac}) are related in a way similar to the \sm\ (\ref{csm}).

\begin{thm}
\label{thm39}
The Schreier array of $\xj$ and $\{\bigy{k}{\rho}{\sigma}\}$ is a 
composition chain of $B$ which is a refinement of the \sm\ of
$\xj$ and $\yk$ given in (\ref{csm}).  Fix $j$ such that $0\le j<\ell$.
Let $J,H$ be groups such that $X_{j-1}^*\le J\le H\le X_j$.  Define 
function $\psi:  H//J\ra\caln(H)//\caln(J)$
by the assignment $\psi:  Jh\mapsto\caln(Jh)$.  The assignment $\psi$
gives a one to one correspondence between all the right cosets 
$Jh$ of $H//J$ and all the right cosets $\caln(Jh)$ of $\caln(H)//\caln(J)$.
Moreover $\psi$ is an $H$-isomorphism giving 
$$
H//J\simeq\caln(H)//\caln(J).
$$

By (v) of Proposition \ref{prop40},
we have $J\lhd H$ \ifof\ $\caln(J)\lhd\caln(H)$.  Thus if 
$J\lhd H$ or $\caln(J)\lhd\caln(H)$, then $\psi$ is an isomorphism
giving $H/J\simeq\caln(H)/\caln(J)$.
In particular let
$$
H=X_{j-1}(X_j\cap\bigy{k}{\rho}{\sigma})
$$
and 
$$
J=X_{j-1}(X_j\cap\bigy{k}{\rho}{\sigma-1}),
$$
where $k>0$ and $\rho\ge -1$, such that $k+\rho<\ell$,
and $0<\sigma\le\sigma_{k+\rho}$.  Then $\psi$ gives an isomorphism
\begin{align}
\label{nmber44}
\frac{X_{j-1}(X_j\cap\bigy{k}{\rho}{\sigma})}
{X_{j-1}(X_j\cap\bigy{k}{\rho}{\sigma-1})}
&\simeq\frac{\caln(X_{j-1}(X_j\cap\bigy{k}{\rho}{\sigma}))}
{\caln(X_{j-1}(X_j\cap\bigy{k}{\rho}{\sigma-1}))} \\
\label{nmber45}
&=\frac{X_j(X_{j+1}\cap\bigy{k-1}{\rho+1}{\sigma})}
{X_j(X_{j+1}\cap\bigy{k-1}{\rho+1}{\sigma-1})}.
\end{align}
This is an isomorphism between terms at the intersection of two lines
with two pages in the Schreier array.
\end{thm}

\begin{prf}
We have already shown that the Schreier array of 
$\xj$ and $\{\bigy{k}{\rho}{\sigma}\}$ is a composition chain
and a refinement of the \sm\ (\ref{csm}).

To show (\ref{nmber44}) we use the isomorphism in (iii)
of Theorem \ref{thm23}.  We only need to verify that
$X_j\cap Y_0\subset J$.  But $X_j\cap Y_0\subset J$
if $Y_0\subset\bigy{k}{\rho}{\sigma-1}$.  Since  
$Y_0=\bigy{1}{-1}{0}$, the inclusion is satisfied if
$k>0$ and $\rho\ge -1$, as assumed. 

We now show (\ref{nmber45}).
We first prove the numerator portion of the equality in (\ref{nmber45}).
We know $X_j^+=X_{j+1}^-$.  From Theorem \ref{thm42} we have
$$
(\bigy{k}{\rho}{\sigma})^+=(\bigy{k-1}{\rho+1}{\sigma})^-.
$$
Therefore
$$
(X_j\cap\bigy{k}{\rho}{\sigma})^+=(X_{j+1}\cap\bigy{k-1}{\rho+1}{\sigma})^-
$$
and 
$$
\caln(X_j\cap\bigy{k}{\rho}{\sigma})=X_0(X_{j+1}\cap\bigy{k-1}{\rho+1}{\sigma}).
$$
From $\caln(GH)=\caln(G)\caln(H)$, we get
$$
\caln(X_{j-1}(X_j\cap\bigy{k}{\rho}{\sigma}))=
X_j(X_{j+1}\cap\bigy{k-1}{\rho+1}{\sigma}).
$$
The proof of the denominator portion of the equality in (\ref{nmber45}) is analogous.
\end{prf}

Theorem \ref{thm39} corresponds to Theorem \ref{thm22a} in the
development of an encoder for $\xj$ and $\yk$.  It follows that 
we can obtain an encoder for $\xj$ and $\{\bigy{k}{\rho}{\sigma}\}$ 
in a similar way as for $\xj$ and $\yk$.  Instead of using 
generators with components from $X_j\cap Y_k$
that are representatives of
$$
\frac{X_{j-1}(X_j\cap Y_k)}{X_{j-1}(X_j\cap Y_{k-1})},
$$
we use generators with components from $X_j\cap\bigy{k}{\rho}{\sigma}$
that are representatives of
$$ 
\frac{X_{j-1}(X_j\cap\bigy{k}{\rho}{\sigma})}
{X_{j-1}(X_j\cap\bigy{k}{\rho}{\sigma-1})}.
$$

\end{document}